\documentclass[aps, twocolumn, pra, superscriptaddress]{revtex4-2}
\usepackage{amsmath}
\usepackage{mathtools}
\usepackage{physics}
\usepackage{datetime}
\newdate{date}{20}{02}{2024}
\usepackage{enumerate}
\usepackage{graphicx} % For figures
\usepackage[hidelinks]{hyperref} % link to refs, eqs, figs, etc. Arxiv complains when we add a particular option, so don't.

\usepackage{amssymb}
\usepackage{comment}
\usepackage{listings}
\usepackage{subfigure} 
\usepackage{booktabs}
\usepackage{makecell}
\usepackage{bm}

\renewcommand{\thefigure}{\textbf{\arabic{figure}}}
\renewcommand{\figurename}{\textbf{Fig.}}

\begin{document}

\title{Observing Dynamical Phases of BCS Superconductors in a Cavity QED Simulator}

\author{Dylan J. Young}
\thanks{These authors contributed equally to this work.}
\affiliation{JILA, NIST, and Department of Physics, University of Colorado, Boulder, CO, USA}
\author{Anjun Chu}
\thanks{These authors contributed equally to this work.}
\affiliation{JILA, NIST, and Department of Physics, University of Colorado, Boulder, CO, USA}
\affiliation{Center for Theory of Quantum Matter, University of Colorado, Boulder, CO, USA}
\author{Eric Yilun Song}
\affiliation{JILA, NIST, and Department of Physics, University of Colorado, Boulder, CO, USA}
\author{Diego Barberena}
\affiliation{JILA, NIST, and Department of Physics, University of Colorado, Boulder, CO, USA}
\affiliation{Center for Theory of Quantum Matter, University of Colorado, Boulder, CO, USA}
\author{David Wellnitz}
\affiliation{JILA, NIST, and Department of Physics, University of Colorado, Boulder, CO, USA}
\affiliation{Center for Theory of Quantum Matter, University of Colorado, Boulder, CO, USA}
\author{Zhijing Niu}
\affiliation{JILA, NIST, and Department of Physics, University of Colorado, Boulder, CO, USA}
\author{Vera M. Sch\"afer}
\affiliation{JILA, NIST, and Department of Physics, University of Colorado, Boulder, CO, USA}
\affiliation{Max-Planck-Institut f\"ur Kernphysik, Saupfercheckweg 1, 69117 Heidelberg, Germany}
\author{Robert J. Lewis-Swan}
\affiliation{Homer L. Dodge Department of Physics and Astronomy, University of Oklahoma, Norman, OK, USA}
\affiliation{Center for Quantum Research and Technology, University of Oklahoma, Norman, OK, USA}
\author{Ana Maria Rey}
\affiliation{JILA, NIST, and Department of Physics, University of Colorado, Boulder, CO, USA}
\affiliation{Center for Theory of Quantum Matter, University of Colorado, Boulder, CO, USA}
\author{James K. Thompson}
\affiliation{JILA, NIST, and Department of Physics, University of Colorado, Boulder, CO, USA}
\usdate
\date{\displaydate{date}}

\begin{abstract}
In conventional Bardeen-Cooper-Schrieffer (BCS) superconductors \cite{bardeen_1957_physrev}, electrons with opposite momenta bind into Cooper pairs due to an attractive interaction mediated by phonons in the material. While superconductivity naturally emerges at thermal equilibrium, it can also emerge out of equilibrium when the system's parameters are abruptly changed \cite{Yuz2,barankov_2006_prl_levitov,Yuz,Gurarie2007,Gurarie2009,Foster2013,yuzbashyan_2015_pra_gurarie}. The resulting out-of-equilibrium phases are predicted to occur in real materials and ultracold fermionic atoms but have not yet all been directly observed.
Here we realise an alternate way to generate the proposed dynamical phases using cavity quantum electrodynamics (cavity QED).
Our system encodes the presence or absence of a Cooper pair in a long-lived electronic transition in $^{88}$Sr atoms coupled to an optical cavity and represents interactions between electrons as photon-mediated interactions through the cavity \cite{lewis-swan_2021_prl_amr,Kelly2022}. To fully explore the phase diagram, we manipulate the ratio between the single-particle dispersion and the interactions after a quench and perform real-time tracking of subsequent dynamics of the superconducting order parameter using non-destructive measurements. We observe regimes where the order parameter decays to zero (phase I) \cite{barankov_2006_prl_levitov,Yuz}, assumes a non-equilibrium steady-state value (phase II) \cite{barankov_2006_prl_levitov,Yuz2}, or exhibits persistent oscillations (phase III) \cite{barankov_2006_prl_levitov,Yuz2}. This opens up exciting prospects for quantum simulation, including the
potential to engineer unconventional superconductors and to probe beyond mean-field effects like the spectral form factor \cite{Stewart2017,Masatoshi2017}, and for increasing coherence time for quantum sensing.

\end{abstract}

\maketitle
\vskip 0.5cm

\section*{Introduction}
Quantum simulation offers a path to understand a broad range of phenomena, from high-temperature superconductivity and correlated quantum magnetism in condensed matter physics \cite{Zhou2021} to quarks and gluons in nuclei and matter under extreme conditions \cite{Shuryak2017}, as well as the black hole information paradox in gravitational physics \cite{Harlow2016}.
A fascinating and promising case is the prethermal dynamical phases \cite{marino_2022_ropp_amr} predicted to emerge from quenches of superconductors and superfluids \cite{Yuz2,barankov_2006_prl_levitov,Yuz,Gurarie2007,Gurarie2009,Foster2013,yuzbashyan_2015_pra_gurarie,VolkovKogan1973,Yuzbashyan2005Lax,Spivak_BCS_2004,Yuzbashyan2008,Foster2014,Collado2023}, systems that feature Cooper pairing of electrons or neutral fermions.
While there has been great progress in pump-probe experiments of superconductors to induce such fast quenches using THz technology, and signs of phases I and II have been observed, the intense pulses couple nonlinearly to the Cooper pairs in the superconductor and complicate a clean observation of the dynamical phases \cite{mansart_2013_pnas_carbone,matsunaga_2013_prl_shimano,matsunaga_2014_science_shimano}.
For these reasons, the realisation of fermionic superfluids in ultracold atomic gases \cite{Randeria:2013kda} has generated great excitement \cite{Yuz2,barankov_2006_prl_levitov,Yuz,Gurarie2007,Gurarie2009,Foster2013,yuzbashyan_2015_pra_gurarie}; however, to date observations have been limited to spectroscopic signatures rather than the full time dynamics \cite{behrle2018higgs}.
In neither system has a systematic scan of the dynamical phase diagram been performed, and in fact phase III has not been observed.

\begin{figure*}
    \includegraphics[keepaspectratio, width=180mm]{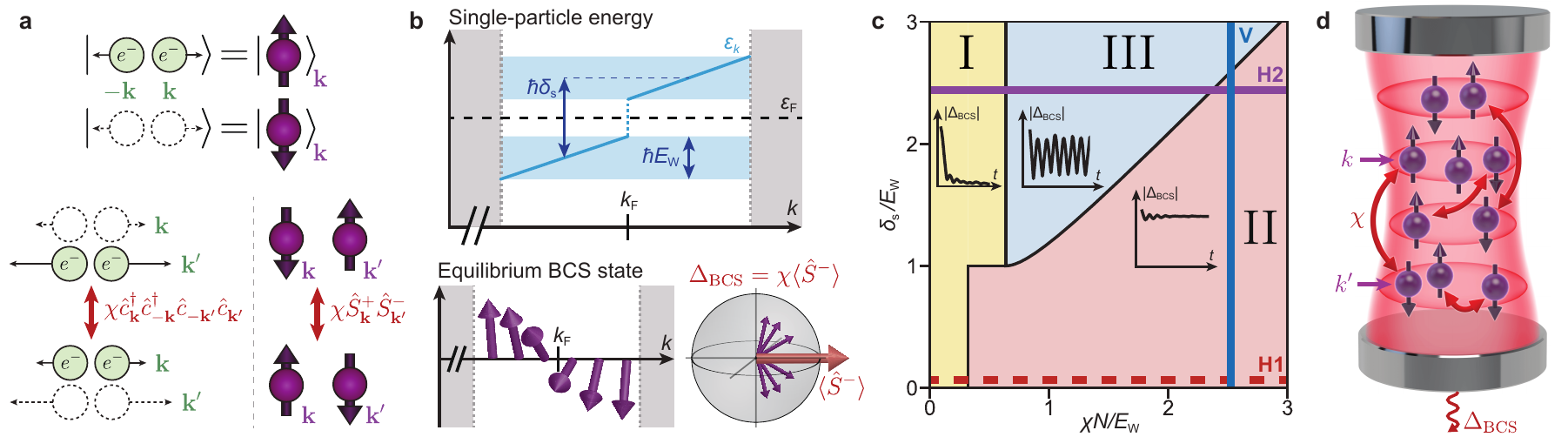}
    \caption{\textbf{Engineering BCS dynamical phases.} \textbf{a}, The Anderson pseudospin mapping encodes the presence and absence of a Cooper pair as the up and down states of a spin-1/2 system, respectively. Under this mapping, the attractive interaction $\chi \hat{c}_\mathbf{k}^\dagger \hat{c}_\mathbf{-k}^\dagger \hat{c}_\mathbf{-k'} \hat{c}_\mathbf{k'}$ between electrons is equivalent to an all-to-all exchange interaction $\chi \hat{S}_\mathbf{k}^{+} \hat{S}_\mathbf{k'}^{-}$ between pseudospins.
    \textbf{b}, Model parameters. The top plot shows the effective dispersion relation near the Fermi surface engineered in our system as a function of parameters $\delta_\mathrm{s}$ and $E_\mathrm{W}$, controlled using AC Stark shifts. The bottom plot visualises the ground state of a BCS superconductor using Anderson pseudospins.
    Near the Fermi momentum, the pseudospins develop a phase-coherent superposition at a scale set by a nonzero BCS pairing gap $\Delta_\mathrm{BCS}$. This gap is self-consistently defined from the spin coherence as shown on the Bloch sphere.
    \textbf{c}, Dynamical phase diagram. The three dynamical phases can be realised by varying parameters $\chi N$, $\delta_\mathrm{s}$, and $E_\mathrm{W}$. Representative dynamics of the BCS order parameter $|\Delta_\mathrm{BCS}|$ for each phase are shown as insets. We explore cut H1 (dashed line) in Fig.~\ref{fig2} using a single ensemble of atoms and cuts V and H2 (solid lines) in Figs.~\ref{fig3} and \ref{fig4} using two separately controlled sub-ensembles.
    \textbf{d}, Cavity QED implementation of the BCS interaction.
    Coupling many strontium atoms to a detuned optical cavity generates infinite-range spin-exchange interactions mediated by a virtual exchange of cavity photons. This interaction also causes a field proportional to $\Delta_\mathrm{BCS}$ to leak out of the cavity, providing a real-time probe of the dynamics.} 
    \label{fig1}
\end{figure*}

Here, we take a step forward towards this challenge by using internal electronic states to encode effective Cooper pairs.
At the heart of this implementation is the Anderson pseudospin mapping \cite{Anderson1958} by which the presence or absence of Cooper pairs in a momentum mode is encoded in a pseudo spin-$1/2$ system. We simulate Anderson pseudospins using a long-lived electronic transition in $^{88}$Sr with interactions between the spins mediated by a high finesse optical cavity. As proposed in Refs.~\cite{lewis-swan_2021_prl_amr,Kelly2022}, the scattering between Cooper pairs in condensed matter systems can be engineered in our system via the exchange of photons through the cavity (see Fig.~\ref{fig1}\textbf{d}). In this way, the dynamics of a collection of interacting spin-$1/2$ systems maps onto the low-energy physics of a superconductor or superfluid.

We probe all three dynamical phases (phases I, II, and III) predicted to exist in BCS superconductors by utilising the high degree of control and flexibility in state initialisation, interaction control, and non-destructive measurements available when coupling long-lived atoms to an optical cavity.
Behaviours intrinsic to phase I (normal phase) and phase II (finite steady-state superconductivity) have previously been observed in spin systems realized in optical cavities \cite{davis_2020_prl_mss,norcia_2018_science_jkt} and in two-level atoms interacting via collisions \cite{allred_2002_prl_romalis,kleine2008,Deutsch:2010ky,smale_2019_science_thywissen}.
We build on this work by clarifying the connection between these dynamical phases from the BCS model and the physics of many-body gap protection in spin systems.
Our results also provide the first demonstration of phase III (a self-generated Floquet phase featuring persistent oscillations of the order parameter), which is predicted to dynamically emerge in superconductors via quenches from weak to strong interactions \cite{barankov_2006_prl_levitov,yuzbashyan_2015_pra_gurarie}. In our system, we instead engineer this phase using flexible control of the single-particle dispersion \cite{lewis-swan_2021_prl_amr,Collado2023}, dynamically resembling the low-energy condition of a BCS superconductor.
For all experiments, we perform real-time tracking of the superconducting order parameter, enabling fast readout of the dynamics.

\section*{Experimental setup and model system}
To realise dynamical phases of the BCS model, we laser cool an ensemble of $N = 10^5-10^6$ $^{88}\mathrm{Sr}$ atoms and trap them inside a $\lambda_{\mathrm{L}} = 813$~nm 1D optical lattice supported by a high-finesse optical cavity. A spin-1/2 system is encoded in the electronic ground state $\ket{\downarrow}=\ket{^{1}\mathrm{S}_0, m_J=0}$ and a long-lived optical excited state $\ket{\uparrow}=\ket{^3\mathrm{P}_1, m_J=0}$. Along this transition, we define spin operators $\hat{S}_{k}^{-} = \ketbra{\downarrow}{\uparrow}_k$ and $\hat{S}_{k}^{z} = (\ketbra{\uparrow}{\uparrow}_{k} - \ketbra{\downarrow}{\downarrow}_{k})/2$ for single atoms with labels $k\in \{1,...,N\}$, as well as the collective lowering operator $\hat{S}^{-} = \sum_{k} \hat{S}_{k}^{-}$ and raising operator $\hat{S}^+=(\hat{S}^{-})^{\dagger}$.

Assuming homogeneous atom-light coupling in the cavity and unitary dynamics, our system can be described by the Hamiltonian
\begin{equation}\label{eq:hamiltonian}
\hat{H} = \hbar\chi \hat{S}^{+} \hat{S}^{-} + \sum_{k} \varepsilon_k \hat{S}_{k}^{z}.
\end{equation}
The first term represents an infinite-range spin-exchange interaction described by a frequency scale $\chi$ \cite{norcia_2018_science_jkt}, realised using the collective coupling between the atomic ensemble and a detuned optical cavity mode. Inhomogeneous atom-light coupling and dissipative processes (including, foremost, single-particle spontaneous decay) are present in the current implementation but do not largely change the qualitative behaviour of the targeted dynamical phases under our experimental conditions (see Methods). Previously, we have characterised this interaction \cite{norcia_2018_science_jkt} and studied collective dynamics by applying an external drive \cite{muniz2020exploring}. In this work, we go beyond the fully collective manifold by engineering a spread in single-particle energies $\varepsilon_k = \hbar \omega_k$ using applied AC Stark shifts $\omega_k$ \cite{baghdad_2023_natphys_long,sauerwein_2023_natphys_brantut}. These shifts form the second term in the Hamiltonian and compete with the spin-exchange interaction. 

Equation~(\ref{eq:hamiltonian}) is the so-called Richardson-Gaudin spin model \cite{Richardson1964A,gaudin1976diagonalization}, which describes the low-energy physics of Bardeen-Cooper-Schrieffer (BCS) superfluids and superconductors using the Anderson pseudospin mapping \cite{Anderson1958}. 
This mapping relates the presence (or absence) of a Cooper pair formed by a pair of electrons with momenta $\pm \mathbf{k}$ to a spin-up (or down) at momentum $\mathbf{k}$, as shown in Fig.~\ref{fig1}\textbf{a}. 
Correspondingly, annihilating a Cooper pair maps to a spin lowering operator by the relation $\hat{S}^-_{\mathbf{k}} \coloneqq \hat{c}_{\mathbf{k}} \hat{c}_{\mathbf{-k}}$, where $\hat{c}_{\pm\mathbf{k}}$ are fermionic annihilation operators. Similarly, the spin operator $2\hat{S}_{\mathbf{k}}^{z}+1 \coloneqq \hat{c}_{\mathbf{k}}^\dagger \hat{c}_{\mathbf{k}}+ \hat{c}_{\mathbf{-k}}^\dagger \hat{c}_{\mathbf{-k}}$ counts the number of electrons with momentum $\mathbf{k}$ or $-\mathbf{k}$. 
Our cavity system therefore manifestly implements a BCS superconductor if one identifies the label $k$ of an atom in the cavity with the momentum $\mathbf{k}$ of the electrons in a Cooper pair.
In this way, the first term in Eq.~(\ref{eq:hamiltonian}) is equivalent to the attractive interaction between electrons in the superconductor, and the second term can be associated with the kinetic energy or dispersion relation of the electrons.
Note that the BCS model, described by Eq.~(\ref{eq:hamiltonian}), only accounts for the zero momentum collective excitations present in conventional superfluids and superconductors \cite{Anderson1958}.

The BCS order parameter in the Anderson mapping is defined by $\Delta_\mathrm{BCS}= \chi \langle \sum_\mathbf{k} 
 \hat{c}_\mathbf{k} \hat{c}_\mathbf{-k} \rangle = \chi \langle \hat{S}^{-} \rangle$, as depicted in Fig.~\ref{fig1}\textbf{b}. 
In equilibrium, it plays the role of the BCS pairing gap, which energetically favours many-body states where the electrons arrange in a coherent superposition between Cooper pairs and holes for states close to the Fermi energy. 
Away from equilibrium, $\Delta_\mathrm{BCS}$ is also predicted to characterise the three dynamical phases (I, II, and III) that arise after quenches in superconductors and superfluids \cite{marino_2022_ropp_amr}. 
Such dynamical phases represent distinct regimes of dynamical behaviour that arise after a sudden perturbation of a control parameter in a closed many-body system. They are described using a time-averaged or steady-state order parameter that demonstrates non-analytic behaviour at the boundary between phases. In particular, the BCS model is predicted to exhibit second-order dynamical phase transitions.

\begin{figure*}
    \includegraphics[keepaspectratio, width=180mm]{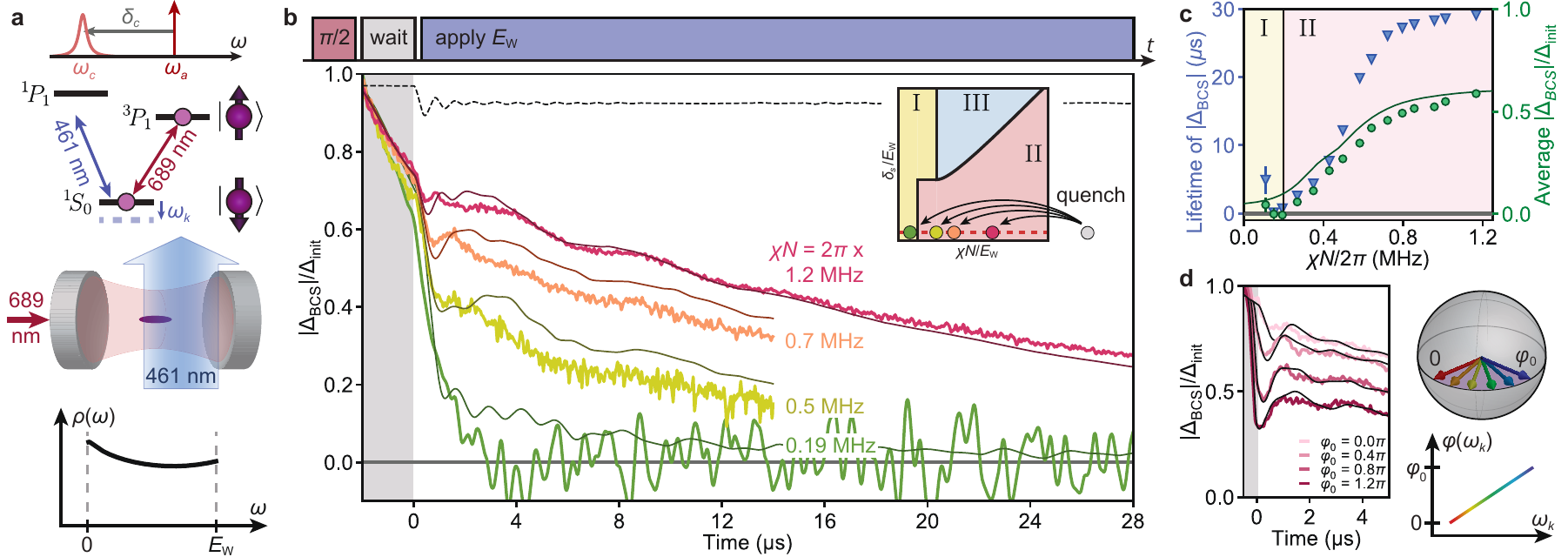}
    \caption{\textbf{Phase I to phase II transition.}
    \textbf{a}, Tuning the single-particle dispersion. We shine an off-resonant 461~nm beam onto the atoms from outside the cavity. This generates a distribution of AC Stark shifts representing a roughly uniform density of states $\rho(\omega)$ (bottom plot).
    \textbf{b}, Probing phase I and phase II. We perform a rapid $\pi/2$ pulse to prepare a highly coherent initial state, wait for $2~\mu$s, quench to a variable $\chi N/E_\mathrm{W}$ with $\delta_\mathrm{s}=0$, and then let the system evolve. The inset shows the explored parameter cut and identifies post-quench $\chi N/ E_\mathrm{W}$ values with coloured dots.
    The main plot shows experimental time traces of $|\Delta_{\mathrm{BCS}}|$ (coloured curves) accompanied by numerical simulations (darker lines). 
    Two curves are extended to demonstrate long-time coherence protection, with the $\chi N/2\pi = 0.19$~MHz trace smoothed for clarity. 
    For $\chi N/2\pi = 1.2$~MHz, we show an ideal simulation neglecting dissipation and motional effects (dashed line), which exhibits transient Higgs oscillations. Hints of these oscillations are present in experimental data with additional damping.
    \textbf{c}, Characterising the phase transition. Blue triangles show the fitted coherence time of $|\Delta_\mathrm{BCS}|$ from $t=1~\mu$s to $30~\mu$s. Green circles show the time-averaged $|\Delta_\mathrm{BCS}|$ between $t=3~\mu$s and $8~\mu$s, with the dark green line representing numerical simulations. In all cases, we identify a phase transition at $\chi N/2\pi = 0.2$~MHz. Error bars in all plots represent the s.e.m.\ of boostrap resamplings on experimental shots.
    \textbf{d}, Varying initial conditions. Before $t=0$, we shine a high-intensity $461$~nm beam within $300$~ns, engineering an initial phase spread $\varphi(\omega_k) \in [0,\varphi_0]$ depicted on the Bloch sphere. The phase $\varphi(\omega_k)$ applied to atom $k$ is proportional to the post-quench frequency shift $\omega_k$.
    Traces represent different $\varphi_0$ and show enhanced oscillations with increasing $\varphi_0$.} \label{fig2}
\end{figure*}

Phase I is characterised by a steady state with a vanishing order parameter $|\Delta_\mathrm{BCS}(t)|\rightarrow0$ at long times. 
Phase II exhibits a steady state with a constant nonzero order parameter $\Delta_\infty \coloneqq \lim_{t\rightarrow\infty} |\Delta_\mathrm{BCS}(t)| > 0$.
Finally, phase III features oscillations in $|\Delta_\mathrm{BCS}(t)|$ that persist to long times, realising a Floquet superfluid despite not being periodically driven \cite{Foster2014,Foster2013,yuzbashyan_2015_pra_gurarie,Gurarie2009}.
The long-time behaviour of these dynamical phases admits a simpler description in terms of the Lax-reduced Hamiltonian, which is an effective Hamiltonian taking the same form of Eq.~(\ref{eq:hamiltonian}) but with rescaled parameters and a reduced number of spins \cite{yuzbashyan_2015_pra_gurarie,marino_2022_ropp_amr}. 
Under this formulation, phases I, II, and III emerge when the Lax-reduced Hamiltonian describes effective zero-spin, one-spin, and two-spin systems respectively. 

Inspired by the Lax-reduced Hamiltonian, and in order to explore all three dynamical phases, we engineer two sub-ensembles of atoms with separate control over energy shifts within each sub-ensemble. For practical convenience, we introduce experimental control in the form of an overall frequency splitting $\delta_\mathrm{s}$ between two sub-ensembles and an effective frequency width $E_\mathrm{W}$ of each sub-ensemble to engineer a tunable dispersion relation $\varepsilon_k$ as in Fig.~\ref{fig1}\textbf{b}. 
Phase I and phase II can also be observed using a single ensemble of atoms as shown in  Fig.~\ref{fig2}.
Both experimental setups can nonetheless be described by a common phase diagram as shown in Fig.~\ref{fig1}\textbf{c}.

We initialise all the atoms in the $\ket{\downarrow}$ state and then apply a coherent $\pi/2$ pulse through the cavity in $100$~ns such that $\Omega \gg \chi N$, where $\Omega$ is the pulse Rabi frequency and $\chi N$ is the characteristic interaction strength for an ensemble of $N$ atoms. This establishes a large BCS order parameter $\Delta_{\mathrm{BCS}}$ on a timescale faster than any other relevant dynamics, mimicking the ground state of a Hamiltonian with an infinite interaction strength $\chi$.
We then quench the system by rapidly turning on $\varepsilon_k$, which sets a finite ratio $\chi N/E_\mathrm{W}$ and a variable $\delta_{\mathrm{s}}/E_{\mathrm{W}}$, 
allowing us to explore the dynamical phase diagram shown in Fig.~\ref{fig1}\textbf{c}.

We measure both the pre- and post-quench dynamics of $|\Delta_\mathrm{BCS}|$ by monitoring light emitted by the atoms into the cavity as a function of time (see Fig.~\ref{fig1}\textbf{d}). This light arises from a superradiance process which is suppressed when the cavity resonance is detuned from the atomic transition frequency by much more than $\kappa$, the cavity power decay linewidth \cite{Weiner_2012_PRA_jkt,bohnet_2013_pra_jkt,norcia_2016_sciadv_jkt}. In this limit, the established cavity field adiabatically follows $\langle \hat{S}^{-} \rangle$, which is proportional to $\Delta_\mathrm{BCS}$. By measuring the leakage of light from the cavity in heterodyne with a local oscillator, we therefore obtain a real-time probe of $\Delta_\mathrm{BCS}$. Importantly, at the chosen detuning this probe is quasi-nondestructive, since only a small fraction of the atoms emit light over relevant timescales. In plots of $|\Delta_\mathrm{BCS}|$ over time, we normalise traces to the initial gap size $\Delta_\mathrm{init}$ measured right after the $\pi/2$ pulse.

\section*{Phase I to phase II}
We probe the phase I to phase II transition by varying the ratio $\chi N/E_\mathrm{W}$ between the interaction strength and the width of the single-particle energy distribution.
As shown in Fig.~\ref{fig2}\textbf{a}, we shine an off-resonant $461$~nm beam onto a single atomic ensemble from the side of the cavity that generates a distribution of AC Stark shifts with a spread $E_\mathrm{W}$.
Careful shaping of the $461$~nm beam allows us to realise a roughly flat density of states (see Methods), resulting in a setup consistent with the $\delta_{\mathrm{s}}=0$ line in Fig.~\ref{fig1}\textbf{c} (see Supplemental Online Material).
After the initial $\pi/2$ pulse, we wait for $2~\mu$s to let transient dynamics settle and then turn on the $461$~nm beam to quench on $E_\mathrm{W}/2\pi = 0.83$~MHz from an initial value $E_\mathrm{W}^{(0)}/2\pi \ll 0.1$~MHz. The beam exhibits a rise time of roughly $50$~ns, much faster than the relevant dynamics.
To scan across the phase diagram in the inset of Fig.~\ref{fig2}\textbf{b}, we vary the interaction strength $\chi N$ between shots by changing the atom number $N$.

\begin{figure*}
    \includegraphics[keepaspectratio, width=180mm]{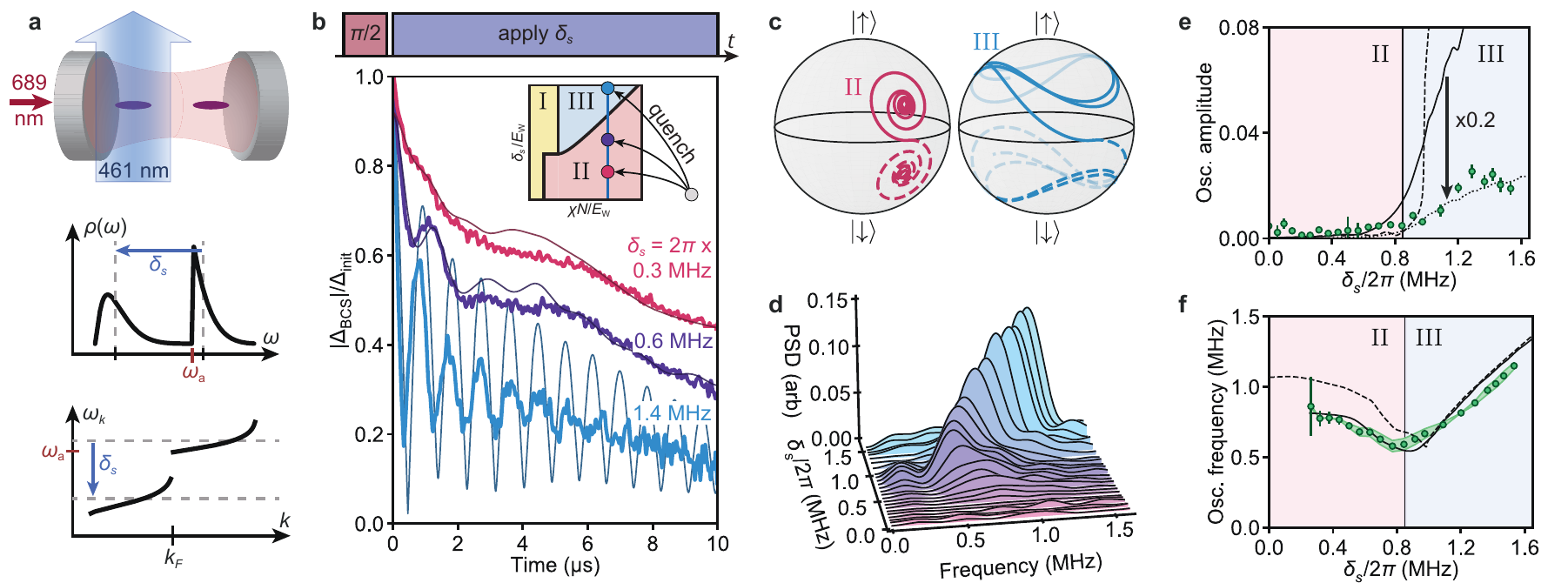}
    \caption{\textbf{Phase II to phase III transition.} \textbf{a}, Engineering a bimodal energy distribution. We prepare two atomic clouds with centres separated by $3$~mm and shine an off-resonant $461$~nm beam centred on one cloud. This generates a density of states $\rho(\omega)$ (middle plot), equivalent to a dispersion relation $\varepsilon_k = \hbar \omega_k$ (bottom plot).
    \textbf{b}, Probing phase II and phase III. We prepare the same initial state as in Fig.~\ref{fig2}\textbf{b} with a $\pi/2$ pulse, quench to a finite $\delta_\mathrm{s}/E_\mathrm{W}$, and then let the system evolve. The inset shows the explored parameter cut and identifies post-quench $\delta_\mathrm{s}/ E_\mathrm{W}$ values with coloured dots.
    As before, coloured traces represent experimental time traces of $|\Delta_{\mathrm{BCS}}|$, and darker lines represent numerical simulations.
    \textbf{c}, Ideal simulations of mean-field trajectories for the two sub-ensembles (solid and dashed curves) in phase II (magenta) and phase III (blue). The trajectories are projected onto the surface of the Bloch sphere for visual clarity.
    \textbf{d}, Fourier response of $|\Delta_{\mathrm{BCS}}|^2$ for different $\delta_\mathrm{s}$, plotted as power spectra of the dynamics from $t=0.5~\mu$s to $4~\mu$s after subtracting slow-moving behaviour.
    \textbf{e}, Average oscillation amplitude between $t=3~\mu$s and $8~\mu$s. For the remaining plots, dashed lines represent ideal simulations (ignoring dissipation or motional effects), and solid dark lines correspond to full simulations. The additional dotted line represents numerical simulations rescaled by $\times 0.2$, plotted to show similar trend behaviour between experimental data and simulations. We identify a phase transition around $\delta_\mathrm{s}/2\pi = 0.85$~MHz.
    \textbf{f}, Oscillation frequency of $|\Delta_{\mathrm{BCS}}|$, measured using power spectra calculated in (\textbf{d}). We correct for systematics inferred from our data analysis and assume this correction has an uncertainty of 100\%, shown by the green band. The phase transition point observed in data in panels (\textbf{e}) and (\textbf{f}) agrees well with simulations.}\label{fig3}
\end{figure*}

As shown in Fig.~\ref{fig2}\textbf{b} and \textbf{c}, we observe two distinct dynamical behaviours corresponding to phases I and II, signalled by the decay rate of $|\Delta_\mathrm{BCS}|$. 
For experiments with sufficiently small $\chi N$, such as $\chi N/2\pi = 0.19$~MHz, $|\Delta_\mathrm{BCS}|$ decays with a $1/e$ coherence time of $0.9 \pm 0.1~\mu$s.
This coherence time is consistent with single-particle dephasing of $\langle \hat{S}^{-} \rangle$ set by the energy spread $\hbar E_\mathrm{W}$ and is nearly constant throughout this regime. 
We identify the fast decay of $|\Delta_\mathrm{BCS}|$ as an experimental signature of phase I. 
For larger interaction strengths, we observe a rapid increase in coherence time up to a maximum of $29~\mu$s when $\chi N/2\pi = 1.2$~MHz; this constitutes an improvement of more than a factor of 30.
We identify this extended coherence time regime as phase II. The residual decay of $|\Delta_\mathrm{BCS}|$ in this regime can be attributed to intrinsic dissipative processes including spontaneous emission, off-resonant superradiant emission, and scattering of 461 nm light \cite{norcia_2016_sciadv_jkt,norcia_2018_science_jkt}, which set a maximum predicted coherence time of $29~\mu$s (see Methods).
All experimental observations (coloured traces) are in good agreement with numerical simulations based on experimental conditions (dark lines---see Methods).

Due to the separation of timescales in the decay of $|\Delta_\mathrm{BCS}|$, we are able to determine the boundary between phase I and phase II in our experiment by calculating the average $|\Delta_\mathrm{BCS}|$ in a time window from $3~\mu$s to $8~\mu$s as a function of $\chi N$ (see Fig.~\ref{fig2}\textbf{c}). In this analysis, phase I features a vanishing average $|\Delta_\mathrm{BCS}|$, while phase II sees a nonzero $|\Delta_\mathrm{BCS}|$ that increases with $\chi N$. The sharp rise of average $|\Delta_\mathrm{BCS}|$ around $\chi N/2\pi = 0.2$~MHz indicates a dynamical phase transition, which agrees with the point predicted by numerical simulations.
In a spin-model picture, the BCS pairing gap corresponds to the energy gap between collective angular momentum states, which exists due to the spin-exchange interaction $\chi \hat{S}^{+} \hat{S}^{-}$ \cite{rey_2008_pra_lukin}. Phase II corresponds to the parameter region where such interactions are sufficiently strong to protect against single-particle dephasing. As a result, the observed transition directly relates to previous experiments exploring coherence protection in other systems \cite{davis_2020_prl_mss, norcia_2018_science_jkt, smale_2019_science_thywissen, Deutsch:2010ky, kleine2008, allred_2002_prl_romalis}.

In BCS superconductors, the excitation of a Higgs mode is predicted to occur in phase II. This mode can be characterised by a collective damped oscillation of the order parameter $|\Delta_\mathrm{BCS}|$ with a characteristic frequency of $2\Delta_\infty$ \cite{yuzbashyan_2015_pra_gurarie}.
We observe hints of Higgs oscillations by comparing the experimental trace of $|\Delta_\mathrm{BCS}|$ at $\chi N/2\pi = 1.2$~MHz (red curve in Fig.~\ref{fig2}\textbf{b}) with the dissipation-free simulation (dashed line in Fig.~\ref{fig2}\textbf{b}) and noticing that the first dip in the experimental trace coincides with the first cycle of Higgs oscillations (see Methods). 
The size of this feature can be increased experimentally by engineering an initial phase spread $\varphi(\omega_k) \in [0,\varphi_0]$ between atoms which is correlated with the post-quench frequency shifts $\omega_k$ of the atoms, as shown in Fig.~\ref{fig2}\textbf{d}.
The initial state with a nonzero opening angle $\varphi_0$ shares qualitative features with the BCS ground state at finite $\chi$ up to a $\pi/2$ rotation on the Bloch sphere \cite{lewis-swan_2021_prl_amr}, in contrast to the initial state mimicking the BCS ground state with infinite $\chi$ in Fig.~\ref{fig2}\textbf{b}.

\section*{Phase II to phase III}
We probe the phase II to phase III transition using a vertical cut through the dynamical phase diagram. 
To realise this, we introduce an energy splitting $\hbar \delta_\mathrm{s}$ between two individually addressable clouds of atoms along the cavity axis using AC Stark shifts from our $461$~nm beam, as shown in Fig.~\ref{fig3}\textbf{a}. In combination with a background energy spread $\hbar E_\mathrm{W}$ associated with lattice shifts (see Methods), this produces a bimodal density of states and a dispersion relation similar to the one proposed in Fig.~\ref{fig1}\textbf{b}. 
As before, we begin the experiment with a highly coherent state and with $\delta_\mathrm{s} = 0$. Then, we quench on a nonzero $\delta_\mathrm{s}$ and let the system evolve. Between shots, we scan $\delta_\mathrm{s}$ while fixing $\chi N/2\pi = 0.9$~MHz and $E_\mathrm{W}/2\pi \approx0.34$~MHz to explore the vertical cut.

The resulting dynamics show a marked change in the dynamical evolution of $|\Delta_\mathrm{BCS}|$ over the scan as shown in Fig.~\ref{fig3}\textbf{b}, which we attribute to a transition between phase II and phase III dynamics. 
For small $\delta_\mathrm{s}$, we either see Higgs-like oscillations which are damped after $3~\mu$s (the trace where $\delta_\mathrm{s}/2\pi = 0.6$~MHz) or, for very small splittings, no oscillations resolvable above the noise floor ($\delta_\mathrm{s}/2\pi = 0.3$~MHz).
We associate this regime with phase II since it overlaps with the previously observed phase II dynamics in parameter space.
For larger $\delta_\mathrm{s}$, curves instead show large-amplitude oscillations that persist for more than $5~\mu$s ($\delta_\mathrm{s}/2\pi = 1.4$~MHz).
We identify the long-lived oscillations in this parameter regime as an experimental signature of phase III.

Intuitively, we can understand the difference between the two phases by identifying the two sub-ensembles of atoms with two Bloch vectors (see Fig.~\ref{fig3}\textbf{c}). In phase II, a finite $\delta_\mathrm{s}$ causes the Bloch vectors to precess in different directions, but the dominant scale $\chi N$ locks them together to form the solid and dashed magenta orbits. In the presence of a finite $E_\mathrm{W}$, the orbits decay, but the Bloch vectors maintain phase coherence. On the other hand, in phase III $\delta_\mathrm{s}$ is large enough that the two Bloch vectors accrue an unbounded relative phase, as in the blue orbits. The presence of interactions locks each sub-ensemble separately against a finite $E_\mathrm{W}$, leading to persistent oscillations.
This effective beating of two large spins in a macroscopic array of spin-1/2 particles is truly an interaction-driven effect since the interactions are strong enough to lock the spins within each sub-ensemble but not strong enough to lock both sub-ensembles together. In our implementation of phase III, the bimodal distribution allows us to dynamically separate the Bloch vectors of the two sub-ensembles, instead of starting with an already split distribution like in weakly interacting BCS ground states featuring a sharp Fermi edge. Despite their qualitative differences, these two situations can be dynamically connected (see Methods).

We can experimentally define a boundary between phase II and phase III using the separation of timescales observed for oscillations in $|\Delta_\mathrm{BCS}|$. Fig.~\ref{fig3}\textbf{e} shows the average oscillation amplitude in a time window from $t = 3~\mu$s to $8~\mu$s. 
In this analysis, we observe a sharp rise in oscillation amplitude at $\delta_\mathrm{s}/2\pi = 0.85$~MHz $\approx \chi N/2\pi$ as we increase $\delta_\mathrm{s}$, which we identify as a dynamical phase transition.
Numerical simulations plotted in Fig.~\ref{fig3}\textbf{e} agree fairly well with data in capturing trend behaviour and estimating the phase transition point. However, we see a discrepancy in the absolute size of the observed and predicted oscillation amplitudes. We attribute this to an extra dephasing mechanism (likely residual motional effects) in our system or other imperfections in the experimental sequence not captured by the theory model.

\begin{figure}
    \centerline{\includegraphics[keepaspectratio, width=89mm]{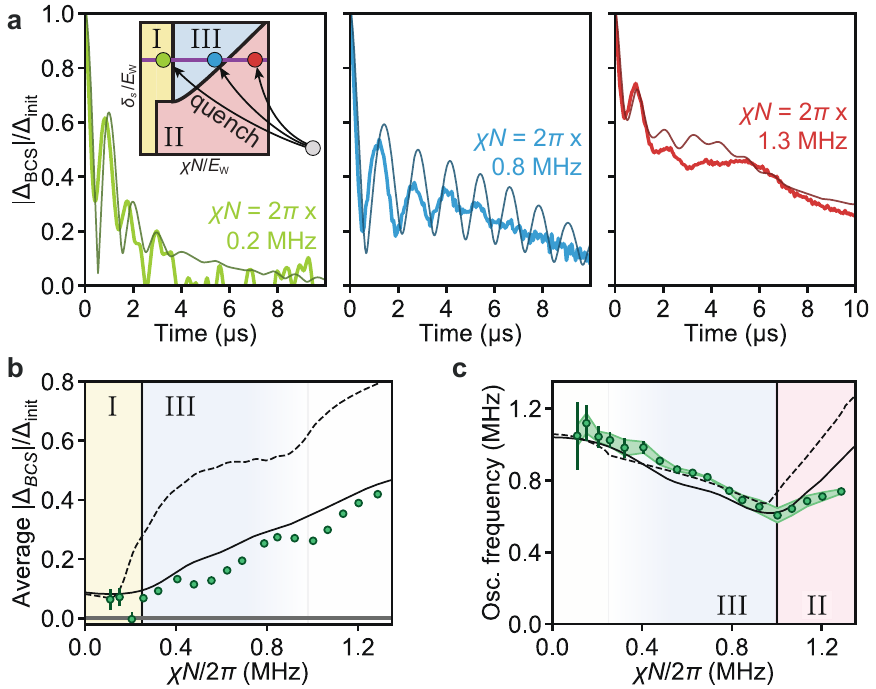}}
    \caption{\textbf{Scan across three dynamical phases.} \textbf{a}, Probing phase I, II and III dynamics using time traces of $|\Delta_\mathrm{BCS}|$. Quenches are performed in the same manner as in Fig.~\ref{fig3}\textbf{b}, except between shots we hold post-quench values of $\delta_\mathrm{s}$ fixed and vary $\chi N$ instead. The inset shows the explored cut through the phase diagram and identifies final $\chi N/ E_\mathrm{W}$ values with green (phase I), blue (phase III), and red (phase II) dots. The $\chi N/2\pi = 0.2$~MHz trace is smoothed for clarity.
    \textbf{b}, Time average of $|\Delta_\mathrm{BCS}|$ in a bin from $t=3~\mu$s to $8~\mu$s vs.\ interaction strength. The experimental data shows signatures of a phase I to phase III transition at $\chi N/2\pi = 0.25$~MHz.
    \textbf{c}, Oscillation frequency of $|\Delta_\mathrm{BCS}|$ vs.\ interaction strength in a bin from $t=0.5~\mu$s to $4~\mu$s. Again, we correct for systematics inferred from our data analysis and assume this correction has an uncertainty of 100\%, shown by the green band. This data identifies a phase III to phase II transition at $\chi N/2\pi = 1.0$~MHz. Experimental data and transitions in both plots are consistent with numerical simulations.}\label{fig4}
\end{figure}

We verify the location of the phase II to phase III transition using the short-time oscillation frequency (from $t=0.5~\mu$s to $4~\mu$s) as an additional experimental signature. As can be seen in the Fourier responses in Fig.~\ref{fig3}\textbf{d} and quantified in Fig.~\ref{fig3}\textbf{f}, the oscillation frequency exhibits a dip vs.\ $\delta_\mathrm{s}$ at the previously-identified phase boundary. This dip is present in roughly the same location for experiment and theory and is expected to coincide with the phase II to phase III transition (see Supplemental Online Material).

\ 
\section*{Scan across three dynamical phases}
Finally, we observe all three dynamical phases in a single cut through parameter space, as shown in Fig.~\ref{fig4}\textbf{a}. We run the same experimental sequence described in Fig.~\ref{fig3}, but instead scan $\chi N$ between shots with $\delta_\mathrm{s}/2\pi = 1.1$~MHz and $E_\mathrm{W}/2\pi = 0.46$~MHz fixed. This allows us to probe phase I, phase III and then phase II by increasing atom number $N$.
Using order parameters established in Figs.~\ref{fig2} and \ref{fig3}, we determine boundaries between the three phases. As shown in Fig.~\ref{fig4}\textbf{b}, the long-time average of $|\Delta_\mathrm{BCS}|$ rises suddenly around $\chi N/2\pi = 0.25$~MHz in both data and simulations. This transition marks the boundary between phase I and phase III. Additionally, at $\chi N/2\pi = 1.0$~MHz we observe a dip in the short-time oscillation frequency of $|\Delta_\mathrm{BCS}|$ (Fig.~\ref{fig4}\textbf{c}), marking a transition between phase III and phase II. For this scan, we do not use the long-time oscillation amplitude as an order parameter due to poor signal-to-noise for smaller values of $\chi N$.

\section*{Conclusion}
The demonstrated capability to emulate dynamical phases of superconductors in optical cavities opens exciting prospects for quantum simulation. For example, it will be interesting to see if our cavity simulator can engineer and probe topological superfluid phases \cite{Foster2013,Black2012,Nandkishore2012,Kiesel2012,Kiesel2013,Fischer2014,Shankar2022} and understand competing superconducting orders \cite{laughlin_1998_prl,Balatsky2006} in a single system, or else enable simulation of superfluidity in phenomena relevant to high energy physics \cite{Schafer2009,Pehlivan2011}.

\bibliographystyle{apsrev4-2}
\bibliography{bibliography}{}

\clearpage
\section*{Methods} 

% Reset figure naming/numbering for Methods
\setcounter{figure}{0}
\renewcommand{\figurename}{\textbf{Extended Data Fig.}}

\subsection{Experimental setup: phase I to phase II transition}
\begin{figure}
    \includegraphics[keepaspectratio, width=89mm]{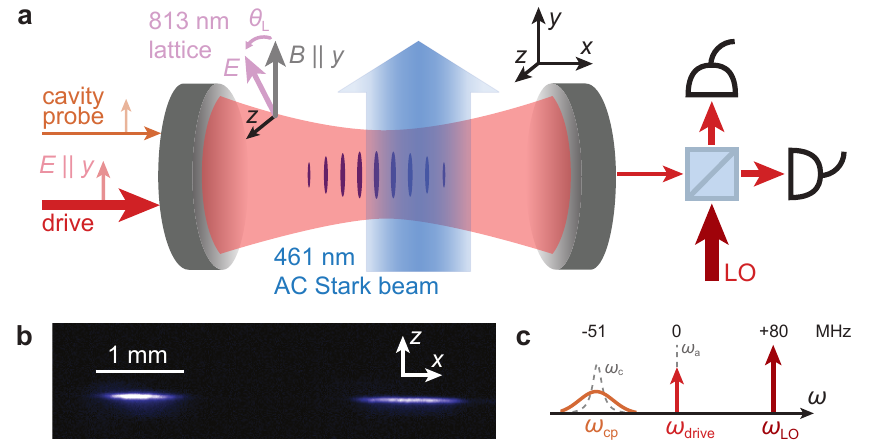}
    \caption{\textbf{Experimental configuration.}
    \textbf{a}, Detailed diagram of the cavity and all relevant beams. A magnetic field along $\hat{y}$ sets the quantisation axis. The 813~nm optical lattice supported by the cavity has a tunable linear polarisation. We drive a $\pi/2$ pulse with a beam polarised along $\hat{y}$ through the cavity, and during the experiment we probe the cavity resonance frequency using a second $\hat{y}$-polarised beam to measure atom number. A 461~nm beam far-detuned from the $\ket{^{1}\!S_{0}} - \ket{^{1}\!P_{1}}$ transition shines on the atoms from the side of the cavity, inducing AC Stark shifts. We probe signals transmitted through the cavity using a balanced heterodyne detector.
    \textbf{b}, Fluorescence image of the two atomic clouds used when scanning through phase III in Figs.~\ref{fig3} and \ref{fig4}.
    \textbf{c}, Frequency landscape of $689$~nm beams. The atomic drive frequency $\omega_\mathrm{drive}$ is resonant with the atomic transition. The cavity probe frequency $\omega_\mathrm{cp}$ is nominally centred with the cavity resonance frequency, $51$~MHz red-detuned from the atomic transition. The local oscillator used in heterodyne detection has frequency $\omega_\mathrm{LO}$ and is 80~MHz blue-detuned from the atomic transition.
    }
    \label{extended_fig1}
\end{figure}

To explore the phase diagram cut in Fig.~\ref{fig2}, we first load $10^5 - 10^6$ $^{88}\mathrm{Sr}$ atoms from a magneto-optical trap into an 813~nm optical lattice supported by a high-finesse optical cavity, similar to previous experiments \cite{norcia_2016_sciadv_jkt,norcia_2018_prx_jkt,norcia_2018_science_jkt,muniz2020exploring}. The resulting atomic cloud has a temperature of roughly $15~\mu$K, resulting in a Gaussian distribution transverse to the cavity axis with standard deviation $\sigma_y = \sigma_z = 16~\mu$m (coordinates defined in Extended Data Fig.~\ref{extended_fig1}\textbf{a}). Further, the cloud is extended over thousands of lattice sites, forming a distribution along the cavity axis with a standard deviation $\sigma_x = 430~\mu$m. We measure an axial trapping frequency of $\omega_x/2\pi = 165~$kHz, giving a Lamb-Dicke parameter of $\eta = 0.17$ for excitation with $689~$nm light. At the measured temperature, $\eta^2(2\bar{n}+1) = 0.11 \ll 1$, placing the atoms in the Lamb-Dicke regime. We set a quantisation axis along $\hat{y}$ with a $2.4$~G magnetic field and tune the lattice polarisation to a ``magic angle'' relative to this axis, such that the differential lattice shift between ground ($\ket{^{1}\!S_0}$) and excited ($\ket{^{3}\!P_1, m_J=0}$) states vanishes \cite{muniz2020exploring}. Using piezoelectric actuators, we stabilise the cavity length to set the closest $\textrm{TEM}_{00}$ resonance to be $51$~MHz red-detuned from the atomic transition.

After loading into the lattice, we initialise the atoms with a $\hat{y}$-polarised drive through the cavity which is nominally resonant with the atomic transition. Because the drive is far off-resonance from the cavity (which has linewidth $\kappa/2\pi = 153$~kHz at $689$~nm), the induced Rabi frequency is somewhat suppressed. Nonetheless, we find that roughly $5$~mW of power before the cavity is sufficient to drive the atoms with a $\pi/2$ pulse in $100$~ns. We allow the atoms to settle for $2~\mu$s in order to distinguish the desired physics from transient dynamics observed after state initialisation, which we attribute to undesired excitation of sideband transitions. We then shine a $461$~nm beam from the side of the cavity along the $\hat{y}$ direction, detuned from the $\ket{^{1}\!S_{0}} - \ket{^{1}\!P_{1}}$ transition by more than $10$~GHz, in order to induce AC Stark shifts on the ground state. The beam has waists $(w_x, w_z) = (1030~\mathrm{\mu m}, 75~\mathrm{\mu m})$ along the $\hat{x}$ and $\hat{z}$ directions at the plane of the atoms, and its centre is displaced from the centre of the atomic cloud by $x_0 = 580~\mathrm{\mu m}$ along the cavity axis. From these dimensions, we calculate an atomic density of states $\rho(\omega)$ as a function of frequency shift which is roughly uniform between $0$ and a maximum shift $\hbar E_\mathrm{W}$. We estimate that for the power and detuning used in this cut, the $461$~nm beam scatters off the atoms with an average rate of $R_\mathrm{sc}/2\pi = 1.3~$kHz, roughly a factor of six smaller than $\gamma/2\pi = 7.5~$kHz, the spontaneous emission rate. Combined with collective emission from the atoms as described in the Readout section of the Methods, these dissipation processes set a maximum predicted coherence time in the system of $29~\mu$s.

\subsection{Experimental setup: cuts through phase III}
For the two cuts through phase III described in Figs.~\ref{fig3} and \ref{fig4}, we load the atoms in two clouds separated by 3~mm, as shown in Extended Data Fig.~\ref{extended_fig1}\textbf{b}. The left cloud has an extent described by standard deviations $(\sigma_x, \sigma_z) = (200~\mathrm{\mu m}, 16~\mathrm{\mu m})$. The right cloud has a similar extent along $\sigma_z$ but is broader along the cavity axis. We tune the lattice polarisation to point along $\hat{z}$, which breaks the magic angle condition and introduces a differential trap depth between ground and excited states of $0.47$~MHz for atoms experiencing peak lattice intensity. Due to their finite temperature, the atoms experience a spread in lattice intensities which leads to an inhomogeneous trap depth. We estimate the induced distribution of energy shifts by assuming the atoms occupy a 2D Gaussian distribution radially with standard deviation $\sigma_y = \sigma_z = 16~\mathrm{\mu m}$, compared to the lattice waist $w_y = w_z = 80~\mathrm{\mu m}$. This produces a peaked distribution equivalent to the narrow peak in Fig.~\ref{fig3}\textbf{a}.

In these experiments, we perform a $\pi/2$ pulse as before and then immediately shine a $461$~nm beam centred on the left (``bright'') atomic cloud. Unlike in the previous cut, we do not wait for transient dynamics to settle after state initialisation, for the sake of simplicity. We do not see major differences between observed and expected behaviour when omitting the wait period. The beam has waists $(w_x, w_z) = (1700~\mathrm{\mu m}, 80~\mathrm{\mu m})$. We install a beam block just before the chamber that clips the beam tail that would otherwise hit the right (``dark'') atomic cloud. The 3~mm separation between clouds is sufficiently large to ensure the beam does not significantly diffract around the beam block. The beam shifts the mean energy of the bright cloud away from that of the dark cloud, introducing a tunable $\delta_\mathrm{s}$. While nominally, we hold $E_\mathrm{W}$ fixed while scanning $\delta_\mathrm{s}$ to explore the phase II to phase III transition, in reality the finite size of the blue beam introduces an additional contribution to $E_\mathrm{W}$ on the bright cloud. As $\delta_\mathrm{s}$ increases, therefore, both the size and shape of the single-particle energy distribution changes. We calculate $E_\mathrm{W}$ in a consistent manner by estimating the standard deviation of the bright cloud distribution and matching the result to a uniform distribution with the same standard deviation (see Supplemental Online Material). In the main text, we report the value of $E_\mathrm{W}$ obtained at the phase transition point for the phase II to phase III transition. As we increase the $461$~nm beam power, the atoms also scatter more blue photons. At the largest applied AC Stark shift, we estimate that the bright cloud experiences a scattering rate of $R_\mathrm{sc}/2\pi = 3.4~$kHz, resulting in lower coherence times for traces with large $\delta_\mathrm{s}$. However, this excess decoherence does not bias our measurements of oscillation amplitude and frequency at times $t \leq 8~\mu$s.

\subsection{Readout}
After initialisation in all experiments, the atomic ensemble establishes a small electric field inside the cavity which adiabatically follows $\langle \hat{S}^{-} \rangle$ \cite{norcia_2018_science_jkt}. Assuming homogeneous atom-light coupling (see next section for modifications due to inhomogeneous coupling), the complex amplitude of the electric field leaking out of the cavity is given by
\begin{equation}
    \alpha_\mathrm{out}(t) = -\frac{g}{\delta_c} \sqrt{\kappa_m} \langle \hat{S}^{-}(t) \rangle,
    \label{eq:sr_field}
\end{equation}
where $\alpha_\mathrm{out}$ has units of $\sqrt{\mathrm{photons}/\mathrm{s}}$. Here, $2g/2\pi = 10.6$~kHz is the single-photon Rabi frequency for an atom maximally coupled to the cavity, $\delta_c/2\pi = (\omega_c-\omega_a)/2\pi = -51$~MHz is the detuning between the cavity resonance frequency $\omega_c$ and the atomic transition frequency $\omega_a$, and $\kappa_m/2\pi = 41~$kHz is the rate at which photons incident on the cavity mirror are transmitted. $\alpha_\mathrm{out}$ is a form of dissipation in the system equivalent to superradiance in a detuned cavity limit. Over the region of parameter space explored in this work, we estimate that the dissipation rate never exceeds $\gamma_{\mathrm{SR}}/2\pi = 2.3$~kHz. We measure the detuned superradiant light as it leaks out of the cavity using balanced heterodyne detection, providing us with a real-time probe of $\langle \hat{S}^{-} \rangle \propto \Delta_\mathrm{BCS}$. In plots of $|\Delta_\mathrm{BCS}|$ in the main text, we calculate the square magnitude of this quantity and average over $400$-$1600$ shots of the experiment, taken within $2$-$10$ minutes. We then perform background subtraction to remove vacuum noise power from the heterodyne signal. Finally, we take a signed square root of the result to return an estimate of $|\Delta_\mathrm{BCS}|$ which averages to zero in the absence of a real signal. This explains why some traces dip below zero despite representing a nonnegative quantity.

Additionally, the cavity experiences a (dispersive) shift in its resonance frequency proportional to the number of atoms. We use this fact to measure atom number by sending a pulsed probe tone through the cavity and measuring the frequency shift using the transmitted light. Since this light is spectrally resolved from the light emitted by the atoms, we are able to measure both signals independently on our heterodyne detector. The different optical frequencies involved in the heterodyne beat are compared in Extended Data Fig.~\ref{extended_fig1}\textbf{c}.

\begin{figure*}
    \includegraphics[keepaspectratio,width=136mm]{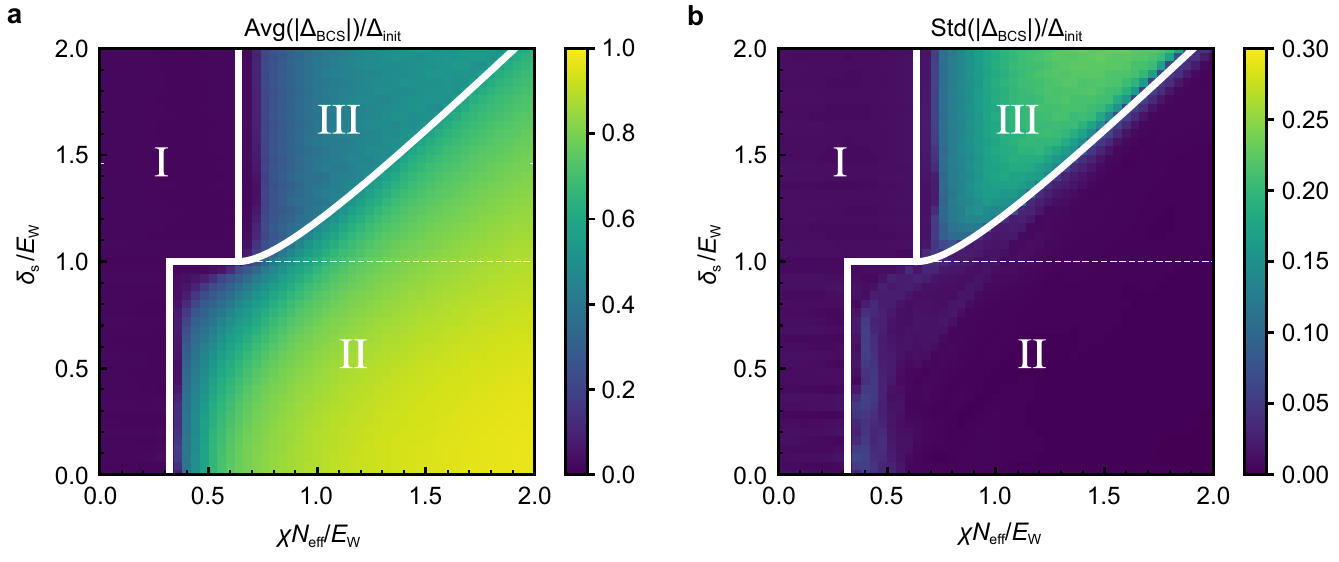}
    \caption{\textbf{Numerical simulation of the dynamical phase diagram based on Eq.~(\ref{eq:inhomo}).} We identify the dynamical phases based on the long-time average (\textbf{a}) and the long-time standard deviation (\textbf{b}) of $|\Delta_{\mathrm{BCS}}(t)|$, normalised by its initial value $\Delta_\mathrm{init}\equiv |\Delta_{\mathrm{BCS}}(0)|$. The white solid lines mark the corresponding dynamical phase boundaries, analytically derived from Eq.~(\ref{eq:hamiltonian}), which agree with the numerical results based on Eq.~(\ref{eq:inhomo}). The white dashed lines mark an extra dynamical phase transition that only exists for Eq.~(\ref{eq:hamiltonian}).}
    \label{extended_fig2}
\end{figure*}

\subsection{Dynamical phase diagram}
The unitary dynamics of our system is modelled by an effective atom-only Hamiltonian, given by
\begin{equation}
    \hat{H}=\hbar\chi\sum_{jk}\zeta_j\zeta_k\hat{S}^{+}_j\hat{S}^{-}_k+\sum_k\varepsilon_k\hat{S}^z_k,
\label{eq:inhomo}
\end{equation}
where $\hat{S}_{k}^{+,-}, \hat{S}_{k}^{x,y,z}$ are the standard spin-1/2 operators on atom $k$. We define $\chi=-g^2\delta_c/(\delta_c^2+\kappa^2/4)$, where $g$ and $\delta_c$ are as defined in the previous section, and $\kappa$ is the cavity linewidth. The spatial dependence of the interaction term is characterised by $\zeta_j=\cos(j\phi)$ with $\phi=\pi\lambda_L/\lambda_c$, which arises because the lattice wavelength $\lambda_L=813$~nm is incommensurate with the cavity wavelength $\lambda_c=689$~nm. In contrast to Eq.~(\ref{eq:hamiltonian}), Eq.~(\ref{eq:inhomo}) becomes non-integrable due to the inhomogeneity in the interaction term. Nevertheless, as shown in Extended Data Fig.~\ref{extended_fig2}, Eq.~(\ref{eq:inhomo}) leads to a similar dynamical phase diagram as Eq.~(\ref{eq:hamiltonian}) if we
\begin{enumerate}[i)]
  \item Use a generalised superconducting order parameter $\Delta_{\mathrm{BCS}}=\chi\sum_k\zeta_k\langle\hat{S}^{-}_k\rangle$;
  \item Interpret the $\pi/2$-pulse as a pulse along the cavity axis under the Hamiltonian $\hat{H}_{\mathrm{drive}}=\hbar\Omega\sum_k\zeta_kS_k^y$ that generates the maximum possible $|\Delta_{\mathrm{BCS}}|$, which occurs when $\Omega t=0.586\pi$;
  \item Replace the atomic number $N$ by an effective atom number $N_{\mathrm{eff}}=N/2$, such that $\chi N_{\mathrm{eff}}$ represents the averaged interaction strength of Eq.~(\ref{eq:inhomo}).
\end{enumerate}

We can still measure the generalised order parameter $\Delta_\mathrm{BCS}$ using the field leaking out of the cavity as in the previous section, since with inhomogeneous coupling the transmitted field takes the form $\alpha_\mathrm{out}(t) = -\tfrac{g}{\delta_c} \sqrt{\kappa_m} \sum_k \zeta_k \langle \hat{S}_k^{-}(t) \rangle \propto \Delta_\mathrm{BCS}$. The dynamical phase diagram in Extended Data Fig.~\ref{extended_fig2} is numerically calculated based on unitary evolution under Eq.~(\ref{eq:inhomo}), with a single-particle dispersion $\varepsilon_k/\hbar$ sampled from a uniform distribution in the frequency range $[-\delta_\mathrm{s}/2-E_\mathrm{W}/2, -\delta_\mathrm{s}/2+E_\mathrm{W}/2]$ and $[\delta_\mathrm{s}/2-E_\mathrm{W}/2, \delta_\mathrm{s}/2+E_\mathrm{W}/2]$. There $\chi N$ corresponds to the averaged interaction strength of Eq.~(\ref{eq:inhomo}). We identify the dynamical phases based on the long-time average of $|\Delta_\mathrm{BCS}|$, given by
\begin{equation}
    \mathrm{Avg}(|\Delta_\mathrm{BCS}|)=\lim_{T\to\infty}\frac{1}{T}\int_0^T|\Delta_\mathrm{BCS}(t)|dt,
\end{equation}
as well as the long-time oscillation amplitude of $|\Delta_\mathrm{BCS}|$. Since the oscillations in $|\Delta_\mathrm{BCS}|$ might deviate from a sinusoidal form, for theoretical simulations it is easier to use the standard deviation as a measure of the oscillation amplitude:
\begin{equation}
    \begin{aligned}
    &\mathrm{Std}(|\Delta_\mathrm{BCS}|)\\
    &=\bigg[\lim_{T\to\infty}\frac{1}{T}\int_0^T\Big(|\Delta_\mathrm{BCS}(t)|-\mathrm{Avg}(|\Delta_\mathrm{BCS}|)\Big)^2dt\bigg]^{1/2}.
    \end{aligned}
\end{equation}
When comparing to experimental data, we measure oscillation amplitude using the Fourier spectrum because technical noise in the experiment contributes to the standard deviation of the time traces (see Fig.~\ref{fig3}\textbf{d}). The dynamical phases can be characterised in theoretical simulations by
\begin{itemize}
    \item Phase I: $\mathrm{Avg}(|\Delta_\mathrm{BCS}|)=0$, $\mathrm{Std}(|\Delta_\mathrm{BCS}|)=0$.
    \item Phase II: $\mathrm{Avg}(|\Delta_\mathrm{BCS}|)>0$, $\mathrm{Std}(|\Delta_\mathrm{BCS}|)=0$.
    \item Phase III: $\mathrm{Avg}(|\Delta_\mathrm{BCS}|)>0$, $\mathrm{Std}(|\Delta_\mathrm{BCS}|)>0$. 
\end{itemize}

\begin{figure*}
    \includegraphics[keepaspectratio,width=136mm]{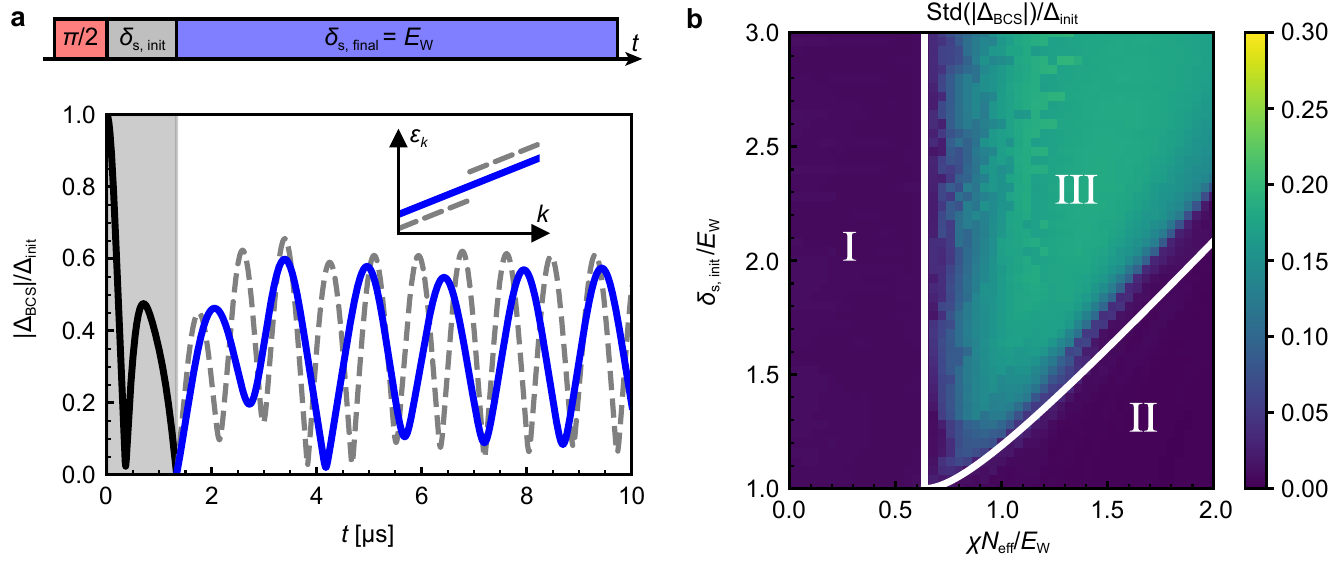}
    \caption{\textbf{Alternative approach for phase III.} \textbf{a}, Simulation of an alternative experimental sequence. As described by the timing sequence at the top, we simulate an experiment that prepares the initial state using a $\pi/2$ pulse, lets the system evolve under a bimodal distribution of single-particle energy (see the inset) until $|\Delta_{\mathrm{BCS}}|$ reaches its minimum value, and then quenches the system back to a continuous distribution of single-particle energies (see the inset). The theoretically predicted time trace of $|\Delta_{\mathrm{BCS}}|$ with $\chi N/E_\mathrm{W}=1.0$ and $\delta_{\mathrm{s,init}}/E_\mathrm{W}=1.6$ is shown at the bottom. The blue (grey dashed) line shows phase III dynamics under a continuous (bimodal) distribution. \textbf{b}, Long-time standard deviation of $|\Delta_{\mathrm{BCS}}(t)|$ after quenching to the continuous distribution shown in \textbf{a}. The white lines are dynamical phase boundaries for bimodal distributions (see Extended Data Fig.~\ref{extended_fig2}). Nearly all the choices of parameter for phase III using bimodal distributions can lead to phase III behaviours after quenching to the continuous distribution.}
    \label{extended_fig3}
\end{figure*}

The dynamical phase boundaries (white solid lines) in Extended Data Fig.~\ref{extended_fig2} are analytically calculated using a Lax analysis applied to Eq.~(\ref{eq:hamiltonian}), similar to the one discussed in \cite{lewis-swan_2021_prl_amr,marino_2022_ropp_amr}, and take the following form (see Supplemental Online Material for a detailed derivation):
\begin{itemize}
    \item Phase I to phase II:
    \begin{equation}
        \begin{gathered}
            \frac{\chi N}{E_\mathrm{W}}=\frac{1}{\pi} \quad \mathrm{with} \quad \frac{\delta_\mathrm{s}}{E_\mathrm{W}}\in [0,1], \\
            \frac{\delta_\mathrm{s}}{E_\mathrm{W}}=1 \quad \mathrm{with} \quad \frac{\chi N}{E_\mathrm{W}}\in \bigg[\frac{1}{\pi},\frac{2}{\pi}\bigg].
        \end{gathered}
    \end{equation}
    \item Phase I to phase III:
    \begin{equation}
        \frac{\chi N}{E_\mathrm{W}}=\frac{2}{\pi} \quad \mathrm{with} \quad \frac{\delta_\mathrm{s}}{E_\mathrm{W}}>1.
    \end{equation}
    \item Phase II to phase III:
    \begin{equation}
        \frac{\delta_s}{E_\mathrm{W}}=\csc\bigg(\frac{E_\mathrm{W}}{\chi N}\bigg) \quad \mathrm{with} \quad \frac{\chi N}{E_\mathrm{W}}>\frac{2}{\pi}.
    \end{equation}
\end{itemize}
The analytical results agree with the numerical simulations for Eq.~(\ref{eq:inhomo}). The only difference is that Eq.~(\ref{eq:hamiltonian}) predicts an extra dynamical phase transition marked by the white dashed line.
The dynamical phase boundaries shown in Fig.~\ref{fig1}\textbf{c} are constructed by the analytical formulas above.

\subsection{Phase III dynamics: the case of a continuous single-particle dispersion}
In this manuscript, we generate phase III using a bimodal single-particle dispersion, represented with idealized assumptions by Fig.~\ref{fig1}\textbf{b} and with actual experimental conditions by Fig.~\ref{fig3}\textbf{a}. Here we show that this experimentally convenient approach generates similar phase III dynamics to the one obtained in the case of a continuous dispersion but with different initial conditions. 

This is done by the protocol shown in Extended Data Fig.~\ref{extended_fig3}\textbf{a}, which uses a bimodal distribution ($\delta_{\mathrm{s,init}}>E_{\mathrm{W}}$) just to generate a state with minimum $|\Delta_{\mathrm{BCS}}|$. At this point the system's dispersion is restored to be continuous by setting $\delta_{\mathrm{s,final}}=E_{\mathrm{W}}$. 
This approach more closely resembles the phase III quench discussed in actual BCS superconductors, where phase III is observed by quenching from a state with weak BCS paring gap $|\Delta_{\mathrm{BCS}}|$ to one with a strong pairing gap \cite{yuzbashyan_2015_pra_gurarie}.
Numerical simulations based on Eq.~(\ref{eq:inhomo}) show that nearly all choices of parameters that lead to phase III using a bimodal distribution also lead to phase III dynamics when quenching to a continuous distribution. The only exception is a small parameter regime close to the boundary between phase III and phase II (see Extended Data Fig.~\ref{extended_fig3}\textbf{b}).
Note that here we use definitions for $\Delta_{\mathrm{BCS}}$, the $\pi/2$ pulse, and $\chi N$ which correspond to Eq.~(\ref{eq:inhomo}), as explained in the previous section.

\subsection{Numerical simulations}
The black dashed lines in Figs.~\ref{fig2}, \ref{fig3}, and \ref{fig4} are computed from unitary evolution under Eq.~(\ref{eq:inhomo}) using a single-particle dispersion $\varepsilon_k$, sampled from the experimentally engineered distribution.

The black solid lines in the same figures are obtained by adding dissipative processes and axial motion to Eq.~(\ref{eq:inhomo}). The system dynamics is described by the following master equation for the density matrix $\hat{\rho}$:
\begin{equation}
 \frac{d\hat{\rho}}{dt}=-\frac{i}{\hbar}[\hat{H},\hat{\rho}]+\mathcal{L}(\hat{L}_c)[\hat{\rho}]+\sum_k\mathcal{L}(\hat{L}_{s,k})[\hat{\rho}]+\sum_k\mathcal{L}(\hat{L}_{\mathrm{el},k})[\hat{\rho}].
\end{equation}
The Lindblad superoperator takes the form $\mathcal{L}(\hat{L})[\hat{\rho}]=\hat{L}\hat{\rho}\hat{L}^{\dag}-\frac{1}{2}(\hat{L}^{\dag}\hat{L}\hat{\rho}+\hat{\rho}\hat{L}^{\dag}\hat{L})$.
Superradiance through the cavity is described by the jump operator
\begin{equation}
 \hat{L}_c=\sqrt{\Gamma}\sum_k\zeta_k\hat{S}^{-}_k,
\end{equation}
where $\Gamma=\chi\kappa/\delta_c$. Spontaneous emission from the atomic excited state is described by the jump operator
\begin{equation}
 \hat{L}_{s,k}=\sqrt{\gamma}\hat{S}^{-}_k,
\end{equation}
where $\gamma/2\pi=7.5~$kHz is the spontaneous emission rate out of $^{3}\!P_1$. Single-particle decoherence is described by the jump operator
\begin{equation}
\hat{L}_{\mathrm{el},k}=\sqrt{2\gamma_{\mathrm{el}}}\hat{S}^{z}_k,
\end{equation}
where $\gamma_{\mathrm{el}}$ is a fitting parameter taking into account free space scattering from the AC Stark shift beam, as well as other decoherence processes in the experiment (see Supplemental Online Material).
These are the dominant dissipative processes in our system. 

The axial trapping frequency of the lattice is $165$~kHz and is therefore smaller than the spin-exchange interaction rate $\chi N$ for most of the experiments. As a consequence, in contrast to the idealised model where atoms are assumed to be frozen, motional processes need to be accounted for, even though they are suppressed in the Lamb-Dicke regime. 
As shown in the Supplemental Online Material, axial motion can lead to a faster damping rate of $|\Delta_{\mathrm{BCS}}|$ oscillations. The predicted dynamical phase boundaries are nevertheless unaffected by the axial motion.

All the numerical simulations are computed using the mean-field approximation, which replaces the operators $\hat{S}^{x,y,z}_k$ by their expectation values $\langle\hat{S}^{x,y,z}_k\rangle$ in the Heisenberg equation of motion. The mean-field treatment of the BCS model is predicted to be exact in the thermodynamic limit due to the infinite-range nature of the interactions \cite{yuzbashyan_2015_pra_gurarie}. The atom number for numerical simulation is set to $5000$ for the ideal conditions and $2000$ for actual experimental conditions. We rescale $\chi$ to match $\chi N$ with experimental values.

\begin{figure}[h]
    \includegraphics[keepaspectratio, width=89mm]{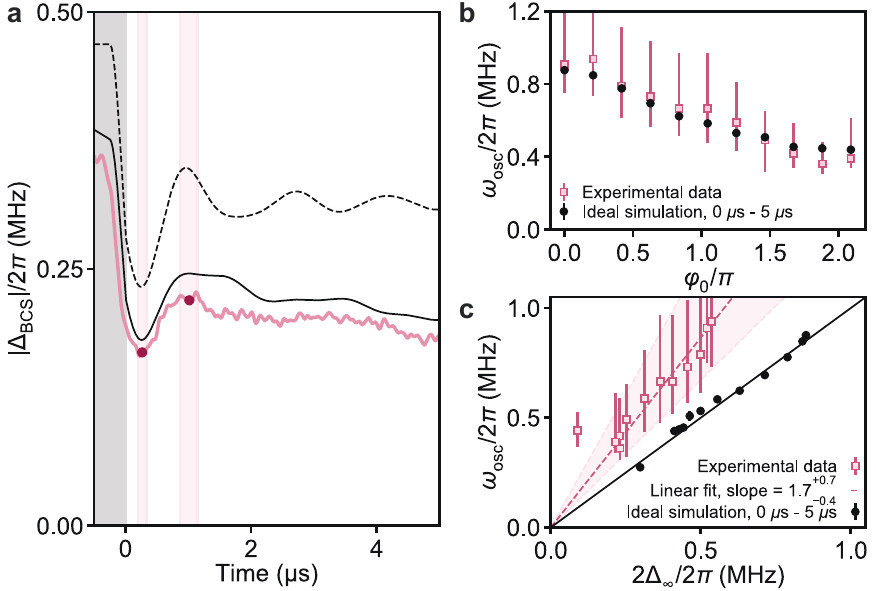}
    \caption{\textbf{Collective scaling in damped phase II oscillations.} \textbf{a}, Time dynamics of $|\Delta_\mathrm{BCS}|$ measured after engineering an initial phase spread over $[0,\varphi_0]$ where $\varphi_0 = 0.8\pi$ as in Fig.~2\textbf{d}, plotted in absolute frequency units (pink trace). The solid black curve represents a numerical simulation of the full system, whereas the dashed curve represents an ideal simulation neglecting dissipation and motional effects. We obtain a crude estimate of oscillation frequency in the experimental data by fitting a trough and peak to smoothed data (after subtracting slow-moving behaviour) within the first couple $\mu$s (magenta points), using these points to infer a half period of oscillation, and with uncertainties determined using a 90\% amplitude threshold (pink bands). \textbf{b}, Comparing oscillation frequency estimates of experimental data (pink squares) with those of ideal simulations (black dots) for different $\varphi_0$. Theory oscillation frequencies are calculated using a Fourier transform from $t = 0~\mu$s to $t = 5~\mu$s. Error bars for experimental data are set by the minimum and maximum frequencies implied by uncertainties in the half period shown in \textbf{a}. The two frequency estimates agree within error bars. \textbf{c}, Collective scaling of oscillation frequency. For each $\varphi_0$ measured in the experiment, we plot the oscillation frequency against the long-time BCS gap $\Delta_\infty$, calculated at $t = 18~\mu$s for ideal simulations and at $t = 3~\mu$s for experimental data. The solid black line is defined by $\omega_\mathrm{osc} = 2\Delta_\infty$, demonstrating the expected scaling for Higgs oscillations. The dashed pink line represents a linear fit to the experimental data. The pink band shows the uncertainty in the slope assuming correlated error in $\omega_\mathrm{osc}$, such that its bounds are defined by linear fits to the data assuming maximum and minimum values for $\omega_\mathrm{osc}$ as defined by the error bars.}
    \label{extended_fig4}
\end{figure}

\subsection{Higgs-like behaviour in short-time phase II dynamics}
When quenching into phase II, we observe highly damped oscillations in $|\Delta_\mathrm{BCS}|$, reminiscent of the Higgs oscillations predicted to arise in this regime of the BCS model. Here, we analyse traces 
from Fig.~\ref{fig2}\textbf{d}, in which we engineer a variable phase spread $\varphi(\omega_k) \in [0,\varphi_0]$ before quenching into phase II, to study this potential connection.

In the BCS model, Higgs oscillations can be characterised by their frequency, which should scale with the long-time BCS order parameter $\Delta_\infty$ as $\omega_\mathrm{osc} = 2\Delta_\infty$ \cite{yuzbashyan_2015_pra_gurarie}. We confirm this scaling in theory by measuring the oscillation frequency from $t = 0~\mu$s to $t = 5~\mu$s in idealised numerical simulations ignoring dissipation and motional effects (black dashed line in Extended Data Fig.~\ref{extended_fig4}\textbf{a}). For different values of the phase spread extent $\varphi_0$, the system reaches its steady state at a different long-time BCS gap $\Delta_\infty$. By parametrically plotting the oscillation frequency vs. $2\Delta_\infty$ as a function of $\varphi_0$ in panel \textbf{c}, we observe the expected scaling.

% Paragraph about measuring experimental frequencies, then panel b.
As discussed in the main text, oscillations in $|\Delta_{\mathrm{BCS}}|$ are consistently smaller and decay more quickly in experiment than in theory. Nonetheless, we obtain a crude estimate of the experimental oscillation frequency by measuring a half period from the first trough and peak of $|\Delta_\mathrm{BCS}(t)|$, as shown in panel \textbf{a}. In panel \textbf{b}, we compare the frequency in experimental data to that of ideal simulations for different $\varphi_0$ and show that the frequencies agree within error bars. This suggests that the transient dynamics observed in $|\Delta_\mathrm{BCS}|$ are related to the Higgs oscillations present in theory.

% Paragraph discussing long-time gap and panel c
Although the experimental oscillation frequency agrees with simulations, the steady-state order parameter $\Delta_\infty$ is much smaller, as can be seen in Extended Data Fig.~\ref{extended_fig4}\textbf{a}. As a result, the measured frequencies scale linearly with $\Delta_\infty$ but with a different prefactor. In panel \textbf{c}, we fit a linear relation of $\omega_\mathrm{osc} = (1.7^{+0.7}_{-0.4})\times 2\Delta_\infty$ to the data, where the slope uncertainty bounds are calculated assuming errors in $\omega_\mathrm{osc}$ are perfectly correlated. Most of the reduction in $\Delta_\infty$ can be captured in theory by considering dissipation and motional effects (solid black trace). We see an additional small difference in $|\Delta_{\mathrm{BCS}}|$ between full numerical simulations and experimental data, which we attribute to drifts in experimental alignments and calibration factors over time. This difference is not apparent in Fig.~\ref{fig2}\textbf{d} because we plot $|\Delta_\mathrm{BCS}|$ in normalised units.

\subsection{Data availability} The datasets generated for this study are available in a Dryad repository with the identifier doi:10.5061/dryad.7h44j100j. \cite{DATA_REPO}

\section*{Acknowledgements}
This material is based upon work supported by the U.S. Department of Energy, Office of Science, National Quantum Information Science Research Centers, Quantum Systems Accelerator. We acknowledge additional funding support from the National Science Foundation under Grant Numbers 1734006 (Physics Frontier Center) and OMA-2016244 (QLCI), NIST, DARPA/ARO W911NF-19-1-0210 and W911NF-16-1-0576, and AFOSR grants FA9550-18-1-0319 and FA9550-19-1-0275. We acknowledge helpful discussions with Emil Yuzbashyan, Victor Gurarie, and Adam Kaufman.
\section*{Author contributions}
D.~J.~Y., E.~Y.~S., Z.~N., V.~M.~S., and J.~K.~T. collected and analysed the experimental data. A.~C., D.~B., D.~W., R.~J.~L.-S., and A.~M.~R. developed the theoretical model. All authors discussed the results and contributed to the preparation of the manuscript.
\section*{Competing interests}
The authors declare no competing interests.
\section*{Additional information} Supplementary Information is available for this paper. Correspondence and requests for materials should be addressed to James K. Thompson (jkt@jila.colorado.edu) or to Ana Maria Rey (arey@jila.colorado.edu). Reprints and permissions information are available at \url{www.nature.com/reprints}.

% Add supplemental to main document to prevent hyperref breaking due to a pdfLaTeX bug
\clearpage
\newpage

\renewcommand{\theequation}{S\arabic{equation}}
\setcounter{equation}{0}
\renewcommand{\thesection}{S\arabic{section}}
\setcounter{section}{0}
\renewcommand{\thetable}{S\arabic{table}}
\setcounter{table}{0}
\renewcommand{\figurename}{\textbf{Fig.}}
\renewcommand{\thefigure}{\textbf{S\arabic{figure}}}
\setcounter{figure}{0}
\renewcommand{\figurename}{\textbf{Fig.}}

\onecolumngrid
\begin{center}
    \textbf{\Large{Supplementary Information}}
\end{center}

\section{Dynamical phase diagram}
In this section, we perform detailed analysis of the dynamical phase diagram shown in Fig.~1\textbf{c} in the main text. We start from analytic calculation in the case of homogeneous couplings, and then generalize to the case of inhomogeneous couplings. Finally we discuss the application of our findings to experimental conditions.

\subsection{Homogeneous model}
First we discuss the dynamical phases for the BCS Hamiltonian with homogeneous couplings,
\begin{equation}
    \hat{H}=\hbar \chi \hat{S}^{+}\hat{S}^{-}+\sum_k\varepsilon_k\hat{S}_k^z.
    \label{eq:bcs}
\end{equation}
We will set $\hbar=1$. As shown in Ref.~\cite{yuzbashyan_2015_pra_gurarie,lewis-swan_2021_prl_amr}, the dynamical phases can be determined using a mean-field Lax vector analysis. The Lax vector is defined as $\vec{L}(u)=L^x(u)\hat{x}+L^y(u)\hat{y}+L^z(u)\hat{z}$ with components,
\begin{equation}
    L^x(u)=\sum_k\frac{S_k^x(0)}{u-\varepsilon_k/2}, \quad L^y(u)=\sum_k\frac{S_k^y(0)}{u-\varepsilon_k/2}, \quad
    L^z(u)=-\frac{1}{\chi}-\sum_i\frac{S_k^z(0)}{u-\varepsilon_k/2},
\end{equation}
where $S^{x,y,z}_k(0)$ are the expectation value of operators $\hat{S}^{x,y,z}_k$ in the initial state. 

Here we consider the initial state as $S^x_k(0)=1/2$, $S^y_k(0)=S^z_k(0)=0$, and $\varepsilon_k$ is chosen from a uniform distribution in the frequency range $[-\delta_\mathrm{s}/2-E_{\mathrm{W}}/2, -\delta_\mathrm{s}/2+E_{\mathrm{W}}/2]$ and $[\delta_\mathrm{s}/2-E_{\mathrm{W}}/2, \delta_\mathrm{s}/2+E_{\mathrm{W}}/2]$. In this case, the mean-field Lax vector takes the following form: 
\begin{equation}
    \begin{gathered}
    \begin{aligned}
    \chi L^x(u)&\approx\frac{\chi N}{2}\bigg[\frac{1}{2E_{\mathrm{W}}}\int_{-\delta_\mathrm{s}/2-E_{\mathrm{W}}/2}^{-\delta_\mathrm{s}/2+E_{\mathrm{W}}/2}\frac{dx}{u-x/2}+\frac{1}{2E_{\mathrm{W}}}\int_{\delta_\mathrm{s}/2-E_{\mathrm{W}}/2}^{\delta_\mathrm{s}/2+E_{\mathrm{W}}/2}\frac{dx}{u-x/2}\bigg]\\
    &=\frac{\chi N}{2E_{\mathrm{W}}}\bigg[\ln\bigg(u+\frac{\delta_\mathrm{s}}{4}+\frac{E_{\mathrm{W}}}{4}\bigg)-\ln\bigg(u+\frac{\delta_\mathrm{s}}{4}-\frac{E_{\mathrm{W}}}{4}\bigg)+\ln\bigg(u-\frac{\delta_\mathrm{s}}{4}+\frac{E_{\mathrm{W}}}{4}\bigg)-\ln\bigg(u-\frac{\delta_\mathrm{s}}{4}-\frac{E_{\mathrm{W}}}{4}\bigg)\bigg],
    \end{aligned}\\
    \chi L^y(u)=0,\\
    \chi L^z(u)=-1.
    \end{gathered}
\end{equation}
Note that $\ln z$ in the complex plane is a multivalued function. Here we take the principal value $\ln z=\ln|z|+i\mathrm{Arg}(z)$, where $\mathrm{Arg}(z)$ is the argument of $z$ restricted in the interval $(-\pi,\pi]$. Directly combining the logarithm functions might lead to moving out of the principal branch.

One can define the dynamical phases based on the number of complex roots of equation $\vec{L}(u)\cdot \vec{L}(u)=0$: Phase I has zero complex roots, phase II has a pair of complex roots, phase III has two pairs of complex roots. Whether the complex roots have non-zero or vanishing real parts could be used for further separation of the phases. In our case, the equation $\vec{L}(u)\cdot \vec{L}(u)=0$ takes the following form,
\begin{equation}
    \frac{\chi N}{2E_{\mathrm{W}}}\bigg[\ln\bigg(u+\frac{\delta_\mathrm{s}}{4}+\frac{E_{\mathrm{W}}}{4}\bigg)-\ln\bigg(u+\frac{\delta_\mathrm{s}}{4}-\frac{E_{\mathrm{W}}}{4}\bigg)+\ln\bigg(u-\frac{\delta_\mathrm{s}}{4}+\frac{E_{\mathrm{W}}}{4}\bigg)-\ln\bigg(u-\frac{\delta_\mathrm{s}}{4}-\frac{E_{\mathrm{W}}}{4}\bigg)\bigg]=\pm i.
    \label{eq:laxequ}
\end{equation}
We find four dynamical phases based on analyzing the roots of Eq.~(\ref{eq:laxequ}):
\begin{itemize}
    \item Phase I: No complex roots, which exist in the regime
    \begin{equation}
        \frac{\delta_\mathrm{s}}{E_{\mathrm{W}}}<1, \quad\frac{\chi N}{E_{\mathrm{W}}}<\frac{1}{\pi} \quad\quad \mathrm{or}\quad\quad \frac{\delta_\mathrm{s}}{E_{\mathrm{W}}}>1, \quad\frac{\chi N}{E_{\mathrm{W}}}<\frac{2}{\pi}.
    \end{equation}

    \item Phase II: A pair of complex roots,
    \begin{equation}
        \frac{u}{E_{\mathrm{W}}}=\pm\frac{i}{4}\bigg[\cot\bigg(\frac{E_{\mathrm{W}}}{\chi N}\bigg)+\sqrt{\csc^2\bigg(\frac{E_{\mathrm{W}}}{\chi N}\bigg)-\frac{\delta^2_s}{E_{\mathrm{W}}^2}}\bigg],
    \end{equation}
    which exist in the regime
    \begin{equation}
        \frac{\delta_\mathrm{s}}{E_{\mathrm{W}}}<1, \quad\frac{\chi N}{E_{\mathrm{W}}}>\frac{1}{\pi}.
    \end{equation}

    \item Phase IIIa: Two pairs of complex roots with vanishing real parts,
    \begin{equation}
        \frac{u_1}{E_{\mathrm{W}}}=\pm\frac{i}{4}\bigg[\cot\bigg(\frac{E_{\mathrm{W}}}{\chi N}\bigg)+\sqrt{\csc^2\bigg(\frac{E_{\mathrm{W}}}{\chi N}\bigg)-\frac{\delta^2_s}{E_{\mathrm{W}}^2}}\bigg], \quad \frac{u_2}{E_{\mathrm{W}}}=\pm\frac{i}{4}\bigg[\cot\bigg(\frac{E_{\mathrm{W}}}{\chi N}\bigg)-\sqrt{\csc^2\bigg(\frac{E_{\mathrm{W}}}{\chi N}\bigg)-\frac{\delta^2_s}{E_{\mathrm{W}}^2}}\bigg],
    \end{equation}
    which exist in the regime
    \begin{equation}
        \frac{\delta_\mathrm{s}}{E_{\mathrm{W}}}>1, \quad\frac{\chi N}{E_{\mathrm{W}}}>\frac{2}{\pi}, \quad \frac{\delta_\mathrm{s}}{E_{\mathrm{W}}}<\csc\bigg(\frac{E_{\mathrm{W}}}{\chi N}\bigg).
    \end{equation}
   In phase IIIa, the order parameter,  $\Delta_\mathrm{BCS}$ oscillates around a non-zero value (non-ZOPA) as pointed out  in Ref.~\cite{Collado2023,lewis-swan_2021_prl_amr}.
    \item Phase IIIb: Two pairs of complex roots with non-zero real parts,
    \begin{equation}
        \frac{u_1}{E_{\mathrm{W}}}=\frac{1}{4}\bigg[\sqrt{\delta'^2_s-\csc^2\bigg(\frac{1}{\chi'N}\bigg)}\pm i\cot\bigg(\frac{1}{\chi'N}\bigg)\bigg], \quad \frac{u_2}{E_{\mathrm{W}}}=\frac{1}{4}\bigg[-\sqrt{\delta'^2_s-\csc^2\bigg(\frac{1}{\chi'N}\bigg)}\pm i\cot\bigg(\frac{1}{\chi'N}\bigg)\bigg],
    \end{equation}
    which exist in the regime
    \begin{equation}
        \frac{\delta_\mathrm{s}}{E_{\mathrm{W}}}>1, \quad\frac{\chi N}{E_{\mathrm{W}}}>\frac{2}{\pi}, \quad \frac{\delta_\mathrm{s}}{E_{\mathrm{W}}}>\csc\bigg(\frac{E_{\mathrm{W}}}{\chi N}\bigg).
    \end{equation}
    In phase IIIb, $\Delta_\mathrm{BCS}$ oscillates with zero order parameter average (ZOPA) as  explained  in Ref.~\cite{Collado2023,lewis-swan_2021_prl_amr}.
\end{itemize} 

\begin{figure*}
  \includegraphics[keepaspectratio, width=170mm]{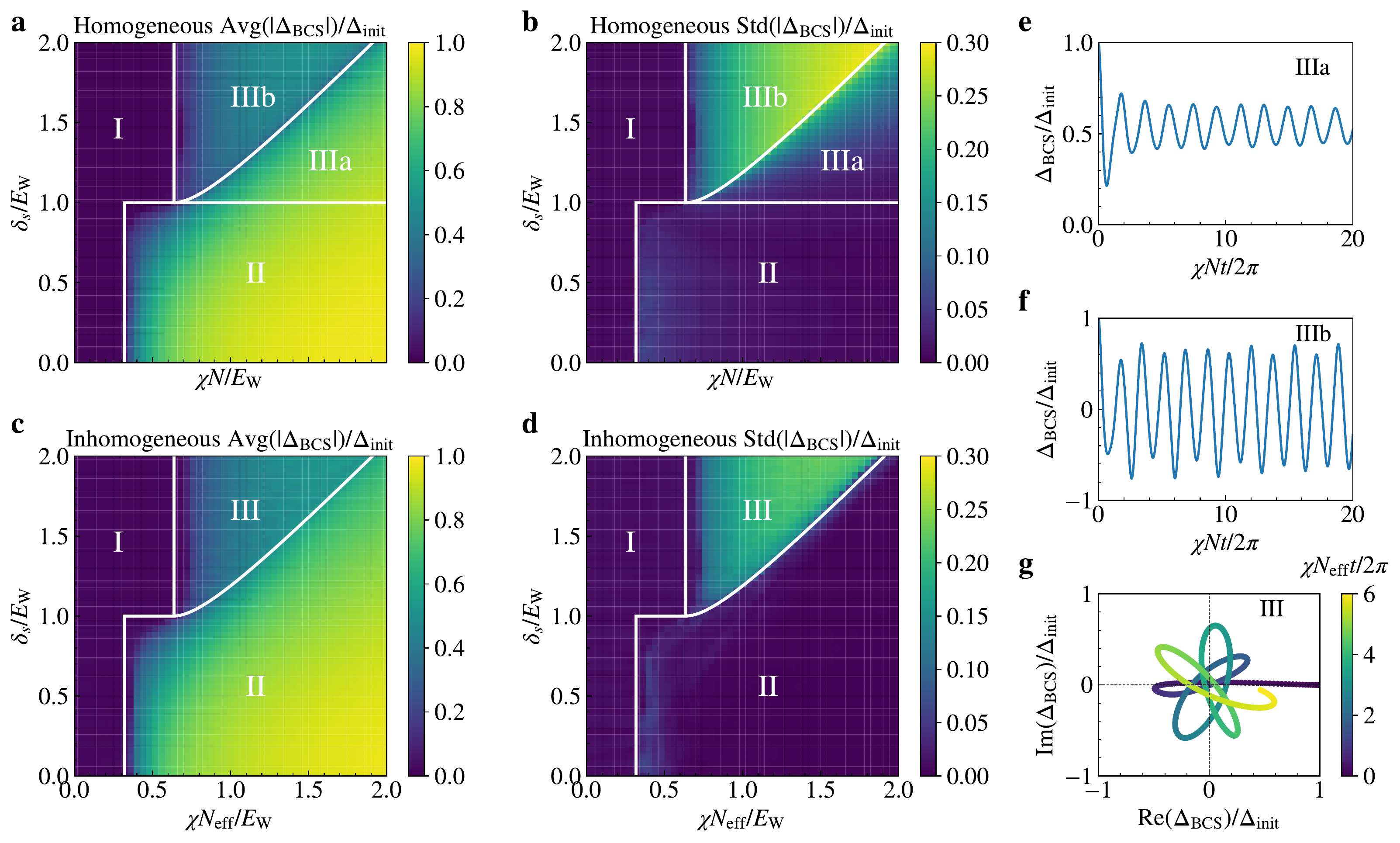}
  \caption{\textbf{Dynamical phase diagrams.} \textbf{a} and \textbf{b}, Dynamical phase diagram of the homogeneous model normalized by $\Delta_{\mathrm{init}}/\chi N=1/2$, where $\Delta_{\mathrm{init}}$ is the initial value of $|\Delta_{\mathrm{BCS}}|$. The white lines are the dynamical critical points derived from the Lax analysis. \textbf{c} and \textbf{d}, Dynamical phase diagram of the inhomogeneous model normalized by $\Delta_{\mathrm{init}}/\chi N_{\mathrm{eff}}=\mathcal{J}_1(\Omega\tau)$. The white lines are the same as the homogeneous model. \textbf{e}, Time evolution of $\Delta_{\mathrm{BCS}}$ at $\delta_\mathrm{s}/E_{\mathrm{W}}=1.1$, $\chi N/E_{\mathrm{W}}=1.0$ under the homogeneous model (phase IIIa). \textbf{f}, Time evolution of $\Delta_{\mathrm{BCS}}$ at $\delta_\mathrm{s}/E_{\mathrm{W}}=1.6$, $\chi N/E_{\mathrm{W}}=1.0$ under the homogeneous model (phase IIIb). \textbf{g}, Time evolution of $\Delta_{\mathrm{BCS}}$ at $\delta_\mathrm{s}/E_{\mathrm{W}}=1.6$, $\chi N/E_{\mathrm{W}}=1.0$ under the inhomogeneous model (phase III).}
  \label{supp0}
\end{figure*}

The dynamical phases derived from the Lax analysis above are supported by numerical evidences, as shown in Fig.~\ref{supp0}\textbf{a} and Fig.~\ref{supp0}\textbf{b}.
We numerically solve the dynamics of $\Delta_\mathrm{BCS}=\chi \langle \hat{S}^{-} \rangle$ under Eq.~(\ref{eq:bcs}) based on mean field approximation, and then identify dynamical phases based on long-time average of $|\Delta_\mathrm{BCS}|$,
\begin{equation}
    \mathrm{Avg}(|\Delta_\mathrm{BCS}|)=\lim_{T\to\infty}\frac{1}{T}\int_0^T|\Delta_\mathrm{BCS}(t)|dt,
\end{equation}
and long-time oscillation amplitude of $|\Delta_\mathrm{BCS}|$. Since the oscillations in $|\Delta_\mathrm{BCS}|$ might deviates from a sinusoidal form, it is easier to use  the standard deviation as a measure of the oscillation amplitude, 
\begin{equation}
    \mathrm{Std}(|\Delta_\mathrm{BCS}|)=\bigg[\lim_{T\to\infty}\frac{1}{T}\int_0^T\Big(|\Delta_\mathrm{BCS}(t)|-\mathrm{Avg}(|\Delta_\mathrm{BCS}|)\Big)^2dt\bigg]^{1/2},
\end{equation}
although experimentally it's better to use the peak of Fourier spectrum to suppress the noise (see Fig.~3\textbf{d} in the main text). The dynamical phases can be characterized by
\begin{itemize}
    \item Phase I: $\mathrm{Avg}(|\Delta_\mathrm{BCS}|)=0$, $\mathrm{Std}(|\Delta_\mathrm{BCS}|)=0.$
    \item Phase II: $\mathrm{Avg}(|\Delta_\mathrm{BCS}|)>0$, $\mathrm{Std}(|\Delta_\mathrm{BCS}|)=0.$
    \item Phase III: $\mathrm{Avg}(|\Delta_\mathrm{BCS}|)>0$, $\mathrm{Std}(|\Delta_\mathrm{BCS}|)>0.$ 
\end{itemize}
Since $\varepsilon_k$ is chosen from a distribution with particle-hole symmetry (symmetric about $0$), $\Delta_\mathrm{BCS}$ becomes a real number in this case. One can further separate phase IIIa and phase IIIb by the behavior of $\Delta_\mathrm{BCS}$ shown in Fig.~\ref{supp0}\textbf{e} and Fig.~\ref{supp0}\textbf{f}.  

\subsection{Inhomogeneous model}
Here we discuss the dynamical phases for the BCS Hamiltonian with inhomogeneous coupling,
\begin{equation}
    \hat{H}=\hbar \chi \sum_{jk}\zeta_j\zeta_k\hat{S}_j^{+}\hat{S}_k^{-}+\sum_k\varepsilon_k\hat{S}_k^z,
    \label{eq:bcsinhomo}
\end{equation}
where $\zeta_k$ is generated by random sampling of $\cos(x)$, with $x$ chosen from a uniform distribution in the interval $[0,2\pi)$. Similar to the homogeneous model, $\varepsilon_k/\hbar$ is still chosen from a uniform distribution in the frequency range $[-\delta_\mathrm{s}/2-E_{\mathrm{W}}/2, -\delta_\mathrm{s}/2+E_{\mathrm{W}}/2]$ and $[\delta_\mathrm{s}/2-E_{\mathrm{W}}/2, \delta_\mathrm{s}/2+E_{\mathrm{W}}/2]$. In this case, we explore the dynamical phases numerically since the Lax analysis is not applicable. As shown in Fig.~\ref{supp1}\textbf{c} and Fig.~\ref{supp1}\textbf{d}, one can obtain similar dynamical phases as the homogeneous model: Phase I remains the same, Phase IIIa merges into Phase II, and Phase IIIb becomes the new Phase III. The phase boundary can be roughly captured by the analytical solution of the homogeneous model. Note that $\chi N_{\mathrm{eff}}$ is the averaged interaction strength in the inhomogeneous case, where $N_{\mathrm{eff}}=N/2$. The superconducting order parameter is defined as $\Delta_{\mathrm{BCS}}=\chi\sum_k\zeta_k\langle \hat{S}_k^{-}\rangle$.
The initial condition is chosen as the maximum $|\Delta_{\mathrm{BCS}}|$ one can achieved by an external drive along the cavity axis, $\hat{H}_{\mathrm{drive}}=\Omega\sum_k\zeta_k\hat{S}_k^y$. Assuming the initial state can be prepared by applying $\hat{H}_{\mathrm{drive}}$ for a time $\tau$, we have
\begin{equation}
    \frac{\Delta_{\mathrm{init}}}{\chi N_{\mathrm{eff}}}\approx \frac{1}{2\pi}\int_0^{2\pi}dx\cos(x)\sin(\Omega\tau\cos(x))=\mathcal{J}_1(\Omega\tau),
\end{equation}
where $\mathcal{J}_n$ is the Bessel function of the first-kind, and the maximum of $\mathcal{J}_1(\Omega\tau)$ can be achieved at $\Omega\tau=0.586\pi$. It is worth to mention that $\Delta_{\mathrm{BCS}}$ is a real number initially, but it becomes a complex number during the time evolution, as shown in Fig.~\ref{supp1}\textbf{g}.

\begin{figure*}
  \includegraphics[keepaspectratio, width=170mm]{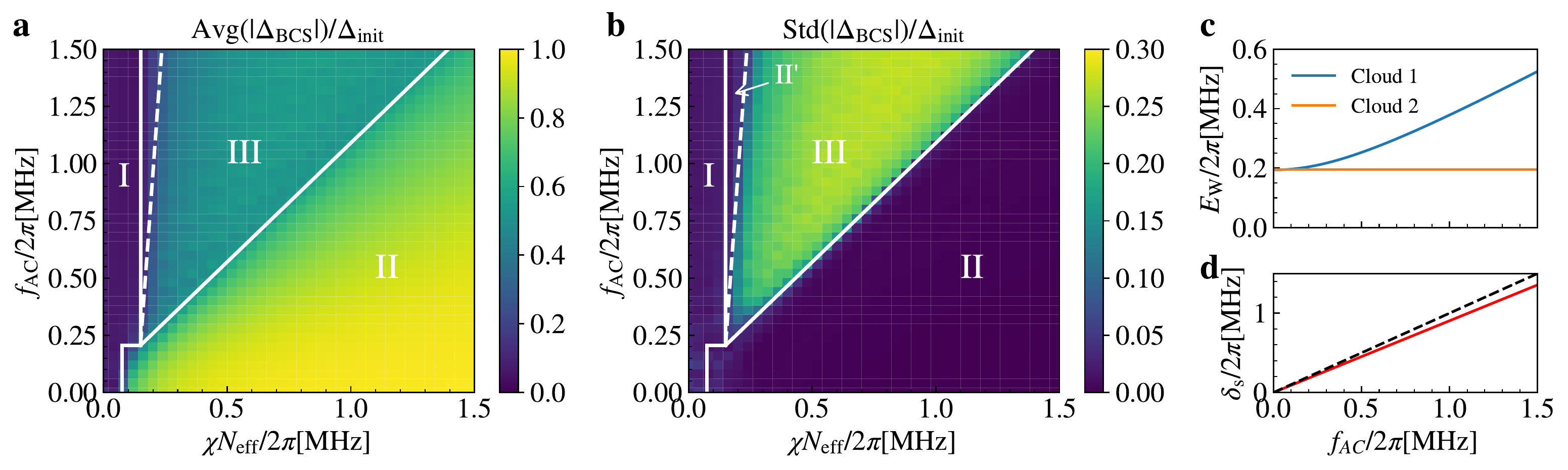}
  \caption{\textbf{Experimental control of dynamical phases.} \textbf{a} and \textbf{b}, Dynamical phase diagram for the experiment with two atomic ensembles, in terms of averaged spin-exchange interaction strength $\chi N_{\mathrm{eff}}$ and peak AC Stark shift $f_{\mathrm{AC}}$. The white lines show the predicted dynamical phase boundaries to guide the eye. The white dashed line marks a small region of phase II' due to the imbalance of $E_\mathrm{W}$ for the two atomic ensembles. \textbf{c}, $E_\mathrm{W}$ as a function of peak AC Stark shift $f_{\mathrm{AC}}$, with AC Stark shift applying to atomic cloud 1. \textbf{d}, $\delta_\mathrm{s}$ as a function of peak AC Stark shift $f_{\mathrm{AC}}$ (red line). The dashed line marks the place where $\delta_\mathrm{s}=f_{\mathrm{AC}}$.}
  \label{supp2}
\end{figure*}

\subsection{Experimental control of dynamical phases}
Here we elaborate on the experimental implementation of the Hamiltonian Eq.~(\ref{eq:bcsinhomo}). As discussed in the previous section, we would like to approximately engineer single-particle energies $\varepsilon_k/\hbar$ sampled from a uniform distribution in the frequency range $[-\delta_\mathrm{s}/2-E_{\mathrm{W}}/2, -\delta_\mathrm{s}/2+E_{\mathrm{W}}/2]$ and $[\delta_\mathrm{s}/2-E_{\mathrm{W}}/2, \delta_\mathrm{s}/2+E_{\mathrm{W}}/2]$. The two different experimental schemes used in the main text to explore the energy distribution are summarized in the following table:

\begin{table}[htbp]
    \centering
    \begin{tabular}{p{3.5cm}p{4.5cm}p{7cm}cc} \toprule
         & \makecell{Description} & \makecell{Approx. $\varepsilon_k/\hbar$} &  & \\ \midrule
    \makecell{Scheme I \\ (Fig.~2, main text)}  &  \makecell{1) Single atomic cloud \\ 2)  AC Stark shift } & \makecell{$[-\tilde{E}_{\mathrm{W}}/2, \tilde{E}_{\mathrm{W}}/2]$} &  & \\ \midrule
    \makecell{Scheme II \\ (Fig.~3, ~4,  main text)} & \makecell{1) Two atomic clouds \\ 2) AC Stark  shift to cloud 1} & \makecell{Cloud 1: $[-\delta_\mathrm{s}/2-E_{\mathrm{W}}/2, -\delta_\mathrm{s}/2+E_{\mathrm{W}}/2]$ \\  Cloud 2: $[+\delta_\mathrm{s}/2-E_{\mathrm{W}}/2, +\delta_\mathrm{s}/2+E_{\mathrm{W}}/2]$} &  & \\ 
    \bottomrule
    \end{tabular}
    %\caption{Caption}
    \label{tab:1}
\end{table}

The first scheme is used to probe the phase I to phase II transition.  We use a single atomic ensemble and apply an AC Stark shift beam with a gradient to approximately engineer $\varepsilon_k/\hbar$ from a uniform distribution $[-\tilde{E}_{\mathrm{W}}/2, \tilde{E}_{\mathrm{W}}/2]$, as discussed in the Methods. As shown in Fig.~2\textbf{a} in the main text, the distribution of atomic frequencies is not exactly uniform, so we calculate the variance of the frequency distribution experimentally. Theoretically we assign a spread $\tilde{E}_{\mathrm{W}}$ such that the uniform distibution over $[-\tilde{E}_{\mathrm{W}}/2, \tilde{E}_{\mathrm{W}}/2]$ matches the measured experimental variance. We use this scheme to probe the dynamical phase diagram at $\delta_\mathrm{s}=0$ (see Fig.~1\textbf{c} in the main text). 

It is worth mentioning that the uniform distribution $[-\tilde{E}_{\mathrm{W}}/2, \tilde{E}_{\mathrm{W}}/2]$ can be interpreted in two different ways: 1) $\delta_\mathrm{s}=0$ and $E_{\mathrm{W}}=\tilde{E}_{\mathrm{W}}$; 2) $\delta_\mathrm{s}=E_{\mathrm{W}}=\tilde{E}_{\mathrm{W}}/2$. Here we prefer the first interpretation $\delta_\mathrm{s}=0$ because in this scheme we only have a single control parameter (the strength of AC Stark shift beam). Additionally, the line $\delta_\mathrm{s}=E_{\mathrm{W}}$ in the dynamical phase diagram has an implication that a small perturbation of $\delta_\mathrm{s}$ can generate a gap in atomic frequency, which is prohibited under this mapping between experimental controls and the model parameters. 

In the second scheme that probes transitions into phase III, we use two atomic ensembles and apply an AC Stark shift beam (peak AC Stark shift $f_{\mathrm{AC}}$) to the first ensemble to generate a frequency splitting $\delta_\mathrm{s}$ between the two ensembles. In contrast to the first scheme, as discussed in the Methods, here we instead use the differential lattice light shifts to engineer a  frequency spread $E_\mathrm{W}$ for each ensemble. As shown in Fig.~3\textbf{a} in the main text, in this case we define $\delta_\mathrm{s}$ as the mean frequency difference between the two ensembles, and $E_\mathrm{W}$ as the width of a uniform distribution generating the same variance.

It is worth mentioning that the Gaussian profile of the AC Stark shift beam leads to an increase in $E_\mathrm{W}$ for the first atomic ensemble, as well as a reduction of the expected splitting of the two ensembles $\delta_\mathrm{s}<f_{\mathrm{AC}}$, as shown in Fig.~\ref{supp2}\textbf{c} and \textbf{d}. 
Using experimental parameters, we get the dynamical phase diagram as depicted in Fig.~\ref{supp2}\textbf{a} and \textbf{b}. 
The imbalance of $E_\mathrm{W}$ for the two atomic ensembles can lead to a small region of phase II' marked by the white dashed line.
This occurs because the spin-exchange interaction is able to lock the ensemble with smaller $E_\mathrm{W}$, while the ensemble with larger $E_{\mathrm{W}}$ remains unlocked, which leads to $|\Delta_{\mathrm{BCS}}|$ approaching a small but nonzero constant value.
In the experiment, due to other dissipative processes and reduced signal-to-noise ratio for small $\chi N$, we do not observe a difference between phase I and phase II'. This is the cause of a small discrepancy between theory and experiment in Fig.~4\textbf{b} in the main text in identifying the position of the phase transition.

\section{Short-time signatures of dynamical phases}
In this section, we discuss the properties of the dynamical phases using short-time observables, since dissipative processes and noise in the experiment lead to difficulties in measuring long-time observables. In the following, we show that phase I can be characterized by the fast decay of $|\Delta_{\mathrm{BCS}}|$, phase II can be characterized by Higgs oscillations.
We further show that the phase II to phase III transition can be captured by the dip in the short-time oscillation frequency of $|\Delta_{\mathrm{BCS}}|$. Finally, we provide an explanation of the frequency dip using an analytical solution of the two-spin BCS model.

\subsection{Phase I: fast decay}
In phase I, the single-particle energy term $\sum_k\varepsilon_k\hat{S}^z_k$ dominates over the spin-exchange interaction. To leading order, one can calculate $|\Delta_{\mathrm{BCS}}|$ in the homogeneous model by dropping the interaction term, which gives
\begin{equation}
    \begin{aligned}
    \frac{|\Delta_{\mathrm{BCS}}|}{\chi N}&\approx \frac{1}{2N}\Big|\sum_ke^{-i\varepsilon_kt/\hbar}\Big|=\frac{1}{2}\bigg|\frac{1}{2E_\mathrm{W}}\int_{-\delta_\mathrm{s}/2-E_\mathrm{W}/2}^{-\delta_\mathrm{s}/2+E_\mathrm{W}/2}e^{-ixt}dx+\frac{1}{2E_\mathrm{W}}\int_{\delta_\mathrm{s}/2-E_\mathrm{W}/2}^{\delta_\mathrm{s}/2+E_\mathrm{W}/2}e^{-ixt}dx\bigg|\\
    &=\frac{1}{2}\bigg|\cos\bigg(\frac{\delta_\mathrm{s}}{2}\bigg)\bigg|\cdot \bigg|\frac{\sin(E_\mathrm{W}t/2)}{E_\mathrm{W}t/2}\bigg|.
    \end{aligned}
\end{equation}
The decay profile of $|\Delta_{\mathrm{BCS}}|$ is set by a sinc function with a $1/e$ coherence time $t$ satisfying $E_\mathrm{W}t/2\pi\approx 0.7$. For the inhomogeneous model a similar fast decay time scale of the order of $E_\mathrm{W}t/2\pi\sim 1$ can be derived. As shown in Fig.~2\textbf{b} in the main text, we observe fast decay of $|\Delta_{\mathrm{BCS}}|$ within $1~\mu$s in phase I. The decay time scale for the other dynamical phases can be more than 10 times longer. 

\subsection{Phase II: Higgs oscillation}
\begin{figure*}
  \includegraphics[keepaspectratio, width=130mm]{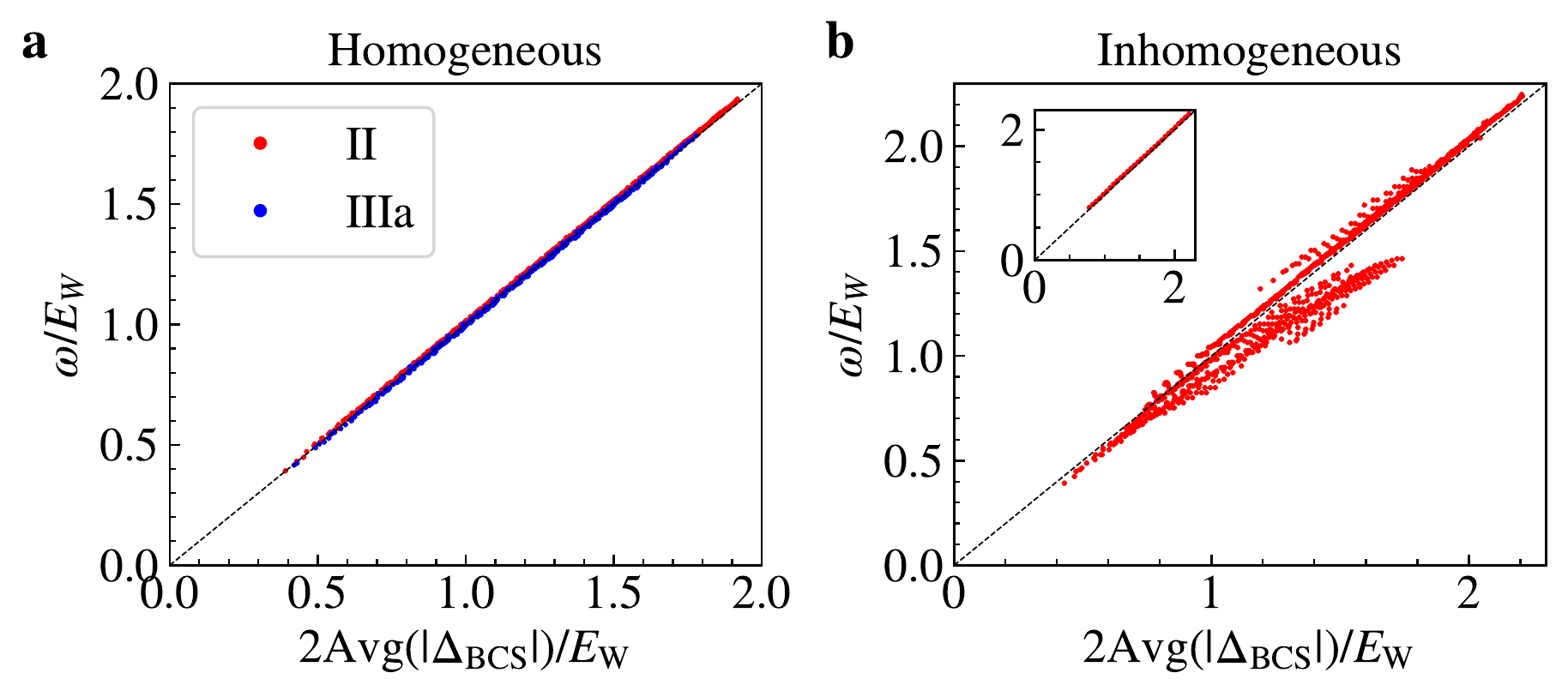}
  \caption{\textbf{Relation between oscillation frequency and averaged order parameter in Higgs oscillations.} \textbf{a}, Homogeneous model where each point is a choice of $(\chi N,\delta_\mathrm{s})$ in phase II (red) and phase IIIa (blue). The dashed line represents $\omega=2\mathrm{Avg}(|\Delta_{\mathrm{BCS}}|)$. \textbf{b}, Inhomogeneous model where each point is a choice of $(\chi N_{\mathrm{eff}},\delta_\mathrm{s})$ in phase II. The inset shows the points with $\delta_\mathrm{s}=0$.}
  \label{supp1}
\end{figure*}

Higgs oscillation, generated by collective excitation of the Higgs mode in BCS superconductor, is characterized by the oscillation of $|\Delta_{\mathrm{BCS}}|$ at frequency $\omega=2\mathrm{Avg}(|\Delta_{\mathrm{BCS}}|)$ \cite{yuzbashyan_2015_pra_gurarie}.
For the homogeneous model (see Fig.~\ref{supp1}\textbf{a}), we numerically confirmed this relation for all the points in phase II and phase IIIa. For the inhomogeneous model (see Fig.~\ref{supp1}\textbf{b}), this relation is approximately satisfied in phase II. In experiment, we observe hints of Higgs oscillation (see Fig.~2 in the main text), which can be ideally described by the inhomogeneous model with $\delta_\mathrm{s}=0$ (see the inset in Fig.~\ref{supp1}\textbf{b}).

\subsection{Transition to phase III: frequency dip}
\begin{figure*}
  \includegraphics[keepaspectratio, width=170mm]{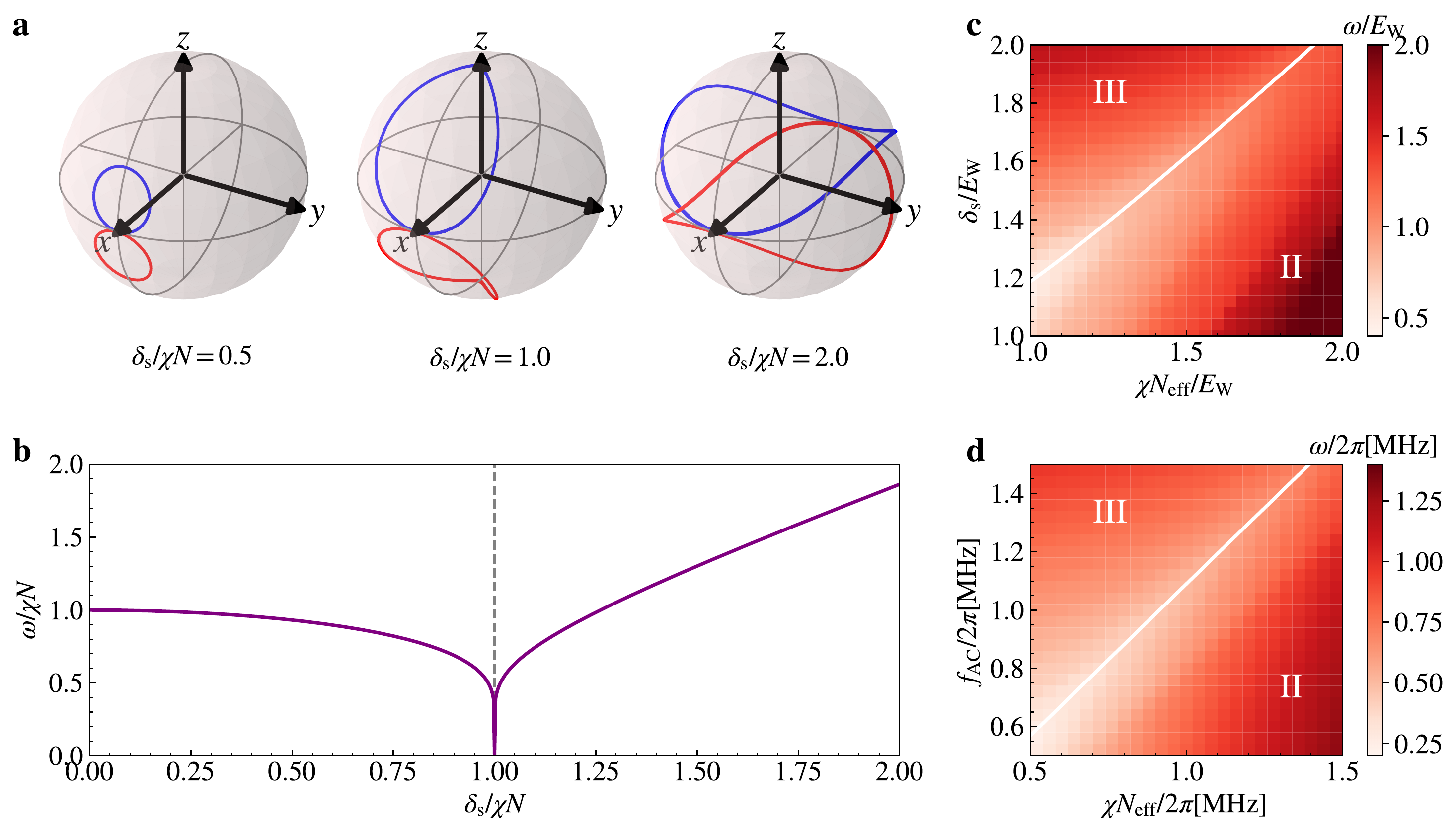}
  \caption{\textbf{Frequency dip as a signature of the phase II to phase III transition.} \textbf{a}, Mean field trajectories of the two large spin model evolving under Eq.~(\ref{eq:twospin}). From left to right, Bloch spheres display trajectories with $\delta_\mathrm{s}/(\chi N) = 0.5,1,$ and 2 respectively. \textbf{b}, Oscillation frequency of $|\Delta_{\mathrm{BCS}}|$ in the two-spin BCS model Eq.~(\ref{eq:twospin}) as a function of $\delta_\mathrm{s}/\chi N$. The frequency dip at $\delta_\mathrm{s}/\chi N=1$ marks the dynamical phase transition point. \textbf{c}, Short-time frequency $\omega$ of the dynamics under inhomogneous atom-light coupling (see Eq.~(3) in the Methods). The white line marks the phase II to phase III transition, the same boundary as shown in Extended Data Fig.~2 from the Methods. \textbf{d}, Short-time frequency $\omega$ of the dynamics using experimental control parameters. The white line marks the phase II to phase III transition and represents the same boundary as in Fig.~\ref{supp2}. The frequency dips match the dynamical critical points for both cases.}
  \label{supp3}
\end{figure*}

In the main text, we discuss a way to understand the phase II to phase III transition by visualising the two atomic ensembles as two large spins. For the inhomogeneous model, phase II exists in the small $\delta_\mathrm{s}$ regime, where the two spins lock to each other and form a single large spin through spin-exchange interactions. In this case the many-body gap protection leads to the damped oscillations observed in phase II. 
Increasing $\delta_\mathrm{s}$ in phase II leads to the reduction of the many-body gap, and hence to a decrease of  the corresponding oscillation frequency. 
Phase III exists in the large $\delta_\mathrm{s}$ regime, where the spin locking occurs separately in each ensemble, and the two large spin are instead precessing around each other, with a rate set by the splitting $\delta_\mathrm{s}$ and the spin-exchange interaction. Increasing $\delta_\mathrm{s}$ in phase III leads to a speed up of the oscillation frequency.
Therefore one expects the existence of a frequency dip separating between phase II and phase III.
Indeed as shown in Fig.~\ref{supp3}\textbf{c} and \textbf{d}, we find good agreement between the frequency dip and the corresponding dynamical critical point. For small $\delta_\mathrm{s}$, the oscillation frequency approaches the Higgs oscillation frequency discussed in the previous subsection. For large $\delta_\mathrm{s}$, the oscillation frequency approaches $\delta_\mathrm{s}$. The reduction of oscillation frequency compared to $\delta_\mathrm{s}$ indicates many-body effects in phase III. It's worth to mention that in contrast to the inhomogeneous model, the frequency dip indicates the phase IIIa to phase IIIb transition for the homogeneous model.

\subsection{Frequency dip in the two-spin BCS model}
Here we use the analytical mean field solution of the
BCS Hamiltonian with two large spins ($S=N/4$ for each spin) to understand the frequency dip discussed above. In this case, the Hamiltonian simplifies to 
\begin{equation}
    \hat{H}/\hbar=\chi \hat{S}^{+}\hat{S}^{-}+\frac{\delta_\mathrm{s}}{2}\hat{S}^z_1-\frac{\delta_\mathrm{s}}{2}\hat{S}^z_2,
    \label{eq:twospin}
\end{equation}
where $\hat{S}^{\pm}=\hat{S}^{\pm}_1+\hat{S}^{\pm}_2$. The mean field equations of motion for the Hamiltonian above can then be written as
\begin{equation}
    \begin{gathered}
        \frac{d}{dt}S^x_1=2\chi S^yS^z_1-\frac{\delta_\mathrm{s}}{2}S^y_1,\quad
        \frac{d}{dt}S^y_1=-2\chi S^xS^z_1+\frac{\delta_\mathrm{s}}{2}S^x_1,\quad
        \frac{d}{dt}S^z_1=-2\chi(S^y_2S^x_1-S^x_2S^y_1),\\
        \frac{d}{dt}S^x_2=2\chi S^yS^z_2+\frac{\delta_\mathrm{s}}{2}S^y_2,\quad
        \frac{d}{dt}S^y_2=-2\chi S^xS^z_2-\frac{\delta_\mathrm{s}}{2}S^x_2,\quad
        \frac{d}{dt}S^z_2=-2\chi(S^y_1S^x_2-S^x_1S^y_2).\\
    \end{gathered}
    \label{eq:meanfield}
\end{equation}
The spin components without the hat represent the expectation value of the corresponding spin operators. 

In the following, we assume an initial state satisfying $S^x_1=S^x_2=N/4$, $S^y_1=S^y_2=S^z_1=S^z_2=0$. The conserved quantities of the two-spin BCS model are the total magnetisation
\begin{equation}
    S^z=S^z_1+S^z_2=0,
    \label{eq:spin}
\end{equation}
the total energy
\begin{equation}
    E/\hbar=\chi S^{+}S^{-}+\frac{\delta_\mathrm{s}}{2}S^z_1-\frac{\delta_\mathrm{s}}{2}S^z_2=\chi\bigg(\frac{N}{2}\bigg)^2,
    \label{eq:energy}
\end{equation}
as well as the spin length of each of the large spins, $(S^x_1)^2+(S^y_1)^2+(S^z_1)^2=(N/4)^2$, $(S^x_2)^2+(S^y_2)^2+(S^z_2)^2=(N/4)^2$. Using these conserved quantities, one can derive from the mean field equations in Eq.~(\ref{eq:meanfield}) an equation of motion for the BCS order parameter, $\Delta_{\mathrm{BCS}}=\chi S^{-}$. To simplify the notation, we define $\Delta\equiv |\Delta_{\mathrm{BCS}}|/\chi N$, i.e. $\Delta^2=S^{+}S^{-}/N^2$. From Eq.~(\ref{eq:spin}) and Eq.~(\ref{eq:energy}), we obtain
\begin{equation}
    \frac{d}{dt}\Delta^2=-\frac{\delta_\mathrm{s}}{\chi N^2}\frac{d}{dt}S^z_1=\frac{2\delta_\mathrm{s}}{N^2}(S^y_2S^x_1-S^x_2S^y_1),
\end{equation}
which leads to
\begin{equation}
    \begin{aligned}
    \frac{d^2}{dt^2}\Delta^2&=\frac{2\delta_\mathrm{s}}{N^2}\bigg(S^x_1\frac{d}{dt}S^y_2+S^y_2\frac{d}{dt}S^x_1-S^y_1\frac{d}{dt}S^x_2-S^x_2\frac{d}{dt}S^y_1\bigg)\\
    &=4\delta_\mathrm{s}\chi \Delta^2S^z_1-\frac{2\delta_\mathrm{s}^2}{N^2}(S^x_1S^x_2+S^y_1S^y_2).
    \end{aligned}
\end{equation}
From the above conserved quantities, we can create the equivalent expressions $\delta_\mathrm{s} S^z_1=-\chi N^2(\Delta^2-1/4)$, $2(S^x_1S^x_2+S^y_1S^y_2)=N^2\Delta^2-2\times(N/4)^2+2(S^z_1)^2$. Plugging these into the equation of motion gives 
\begin{equation}
    \frac{d^2}{dt^2}\Delta^2=-6(\chi N)^2(\Delta^2)^2+\Big(2(\chi N)^2-\delta_\mathrm{s}^2\Big)\Delta^2+\frac{\delta_\mathrm{s}^2-(\chi N)^2}{8}.
\end{equation}
The equation above can be further simplified to
\begin{equation}
    \frac{1}{2}\bigg(\frac{d}{dt}\Delta\bigg)^2+V(\Delta)=0,
    \label{eq:gap}
\end{equation}
where
\begin{equation}
    V(\Delta)=\frac{1}{2}(\chi N)^2\bigg(\Delta^2-\frac{1}{4}\bigg)\bigg(\Delta^2-\frac{1-(\delta_\mathrm{s}/\chi N)^2}{4}\bigg),
\end{equation}
with an initial condition $\Delta=1/2$. Eq.~(\ref{eq:gap}) can be understood as a classical particle with position $\Delta$ oscillating in the potential $V(\Delta)$. For $\delta_\mathrm{s}<\chi N$, we find $\Delta$ oscillating between $\Delta_{\mathrm{max}}=1/2$ and $\Delta_{\mathrm{min}}=\sqrt{1-(\delta_\mathrm{s}/\chi N)^2}/2$. 
This is equivalent to phase II in the cases of many spins with inhomogeneous atom-light couplings, because all the oscillations damp in the large $\chi N$ limit.
For $\delta_\mathrm{s}>\chi N$, we find $\Delta$ oscillating between $\Delta_{\mathrm{max}}=1/2$ and $\Delta_{\mathrm{min}}=0$, since the definition of $\Delta$ requires $\Delta\geq 0$. 
This is equivalent to phase III in the cases of many spins because the phase connects to single-particle oscillations in the large $\delta_\mathrm{s}$ limit.
Therefore, a dynamical phase transition occurs at $\delta_\mathrm{s}/\chi N=1$, which is equivalent to the phase II to phase III transition in the many-spin system.

The analytical solution of Eq.~(\ref{eq:gap}) can be written in terms of Jacobian elliptic funtions $\mathrm{dn}$ and $\mathrm{cn}$:
\begin{equation}
    \Delta(t)=\begin{cases}
        \displaystyle\;\frac{1}{2}\mathrm{dn}\bigg(\frac{1}{2}\chi Nt\bigg|(\delta_\mathrm{s}/\chi N)^2\bigg) \quad \mathrm{if}\;\delta_\mathrm{s}<\chi N\\
        \\
        \displaystyle\;\frac{1}{2}\bigg|\mathrm{cn}\bigg(\frac{1}{2}\delta_\mathrm{s} t\bigg|(\chi N/\delta_\mathrm{s})^2\bigg)\bigg| \quad \mathrm{if}\;\delta_\mathrm{s}>\chi N
    \end{cases}.
    \label{eq:tra}
\end{equation}
The frequency of $\Delta(t)$ can be written in terms of the complete elliptic integral of the first kind $K(k^2)$:
\begin{equation}
    \frac{\omega}{\chi N}=\begin{cases}
    \displaystyle\; \frac{\pi}{2K\Big((\delta_\mathrm{s}/\chi N)^2\Big)} \quad \mathrm{if}\;\delta_\mathrm{s}<\chi N\\
    \\
    \displaystyle\; \frac{\delta_\mathrm{s}}{\chi N}\frac{\pi}{2K\Big((\chi N/\delta_\mathrm{s})^2\Big)} \quad \mathrm{if}\;\delta_\mathrm{s}>\chi N\\
    \end{cases}.
    \label{eq:freq}
\end{equation}
The mean-field trajectories on the Bloch sphere are shown in Fig.~\ref{supp3}\textbf{a}, and the oscillation frequency Eq.~(\ref{eq:freq}) is shown in Fig.~\ref{supp3}\textbf{b}. The dynamical phase transition can also be understood from the mean field trajectories. For $\delta_\mathrm{s}<\chi N$, the two large spins lock to each other and oscillate near the $x$ axis of the Bloch sphere. For $\delta_\mathrm{s}>\chi N$, the two large spins are unlocked and precess around the whole Bloch sphere. Near the dynamical critical point, the mean field trajectories are close to the north pole or south pole of the Bloch sphere, which leads to a slow down of the oscillations because they approach stable fixed points of the Hamiltonian.

\section{Axial Motion}
In this section, we elaborate on how to take into account axial motion present in the experimental system. Similar discussions can be found in Ref.~\cite{muniz2020exploring}.
We start with the one-dimensional Hamiltonian of our cavity QED system with two internal atomic levels ($|\negmedspace\uparrow\rangle$ and $|\negmedspace\downarrow\rangle$), given by
\begin{equation}
    \begin{aligned}
    \hat{H}&=\sum_{\sigma=\{\uparrow,\downarrow\}}\int dx\;\hat{\psi}^{\dag}_{\sigma}(x)\bigg[\frac{\hat{p}^2}{2M}+V_{0}\sin^2(k_Lx)\bigg]\hat{\psi}_{\sigma}(x)+\int dx\;\hat{\psi}^{\dag}_{\uparrow}(x)\bigg[\hbar\omega_{0}+U_{ac}(x)\bigg]\hat{\psi}_{\uparrow}(x)\\
    &+\hbar g_c\int dx\;\cos(k_cx)\bigg[\hat{\psi}^{\dag}_{\uparrow}(x)\hat{\psi}_{\downarrow}(x)\hat{a}+\hat{a}^{\dag}\hat{\psi}^{\dag}_{\downarrow}(x)\hat{\psi}_{\uparrow}(x)\bigg]+\hbar\omega_c\hat{a}^{\dag}\hat{a},\\
    \end{aligned}
\end{equation}
where $k_L=2\pi/\lambda_L$ is the wavenumber of the lattice beams ($\lambda_L=813$nm), $k_c$ is the wavenumber of the cavity mode ($\lambda_c=689$nm), $\omega_{0}$ is the atomic transition frequency between $|\negmedspace\uparrow\rangle$ and $|\negmedspace\downarrow\rangle$ states, $U_{ac}(x)$ is the AC Stark shift applied to the atoms (including the differential light shift from the lattice beams and the transverse AC Stark shift beam), and $\omega_c$ is the frequency of cavity resonance.
%$\omega_p$ is the frequency of pump laser.

Since the atoms are trapped in an optical lattice with lattice depth on the order of $10^3E_R$, we can approximate each lattice site as an harmonic trap with axial trapping frequency $\hbar\omega_{T}=\sqrt{4V_{0}E_R}$, where $E_R=\hbar^2k_L^2/2M$ is the lattice recoil energy. We also ignore tunnelling processes between lattice sites. In this case, one can expand the atomic field operator in terms of lattice site index $j$ and harmonic oscillator levels $n$:
\begin{equation}
    \hat{\psi}_{\sigma}(x)=\sum_{jn}\hat{c}_{jn,\sigma}\phi_n(x-ja_L).
\end{equation}
Here, $a_L=\lambda_L/2$ is the lattice spacing, and $\phi_n$ is the harmonic oscillator wave function for mode $n$, given by
\begin{equation}
    \phi_n(x)=\frac{1}{\sqrt{2^nn!}}\bigg(\frac{M\omega_{T}}{\pi\hbar}\bigg)^{1/4}\mathrm{e}^{-M\omega_{T} x^2/2\hbar}H_n\bigg(\sqrt{\frac{M\omega_{T}}{\hbar}}x\bigg)
\end{equation}
where $H_n(x)$ are the Hermite polynomials. Plugging this expansion into the Hamiltonian and transforming to the rotating frame of the atoms, we obtain 
\begin{equation}
\hat{H}/\hbar=\sum_{jn\sigma}n\omega_T\hat{c}^{\dag}_{jn,\sigma}\hat{c}_{jn,\sigma}+\sum_{jn}\varepsilon_{jn}\hat{c}^{\dag}_{jn,\uparrow}\hat{c}_{jn,\uparrow}+g_c\sum_{jnm}\zeta_j^{nm}(\hat{c}^{\dag}_{jn,\uparrow}\hat{c}_{jm,\downarrow}\hat{a}+\hat{a}^{\dag}\hat{c}^{\dag}_{jm,\downarrow}\hat{c}_{jn,\uparrow})+\delta_c\hat{a}^{\dag}\hat{a}
\end{equation}
where $\delta_c=\omega_c-\omega_a$. For simplicity, we assume $U_{ac}(x)$ is either small or slowly varying in space and thus does not change the trap geometry. This term gives rise to an inhomogeneous transition frequency $\varepsilon_{jn}=\int dx U_{ac}(x)[\phi_n(x-ja_L)]^2/\hbar$. We calculate $\zeta_j^{nm}$ in the following way:
\begin{equation}
    \begin{aligned}
    \zeta_j^{nm}&=\int dx\;\cos(k_cx)\phi_n(x-ja_L)\phi_m(x-ja_L)=\int dx\;\cos(k_cx+k_cja_L)\phi_n(x)\phi_m(x)\\
    &=\cos(j\varphi)\int dx\;\cos(k_cx)\phi_n(x)\phi_m(x)-\sin(j\varphi)\int dx\;\sin(k_cx)\phi_n(x)\phi_m(x)\\
    &=\cos(j\varphi)\,\mathrm{Re}\bigg[(i\eta)^s e^{-\eta^2/2}\sqrt{\frac{n_{<}!}{n_{>}!}}L^s_{n_<}(\eta^2)\bigg]-\sin(j\varphi)\,\mathrm{Im}\bigg[(i\eta)^s e^{-\eta^2/2}\sqrt{\frac{n_{<}!}{n_{>}!}}L^s_{n_<}(\eta^2)\bigg].
    \end{aligned}
\end{equation}
where $\varphi=\pi k_L/k_c$, $s=|n-m|$, $n_{<}=\min(n,m)$, $n_{>}=\max(n,m)$, $L^{\alpha}_n(x)$ are the generalised Laguerre polynomials, and $\eta=k_c\sqrt{\hbar/2M\omega_T}$ is the Lamb-Dicke parameter. In our case $\omega_T/2\pi=165$~kHz, implying that $\eta=0.17$. This places us in the Lamb-Dicke regime where $\zeta^{nm}_j$ is negligible for $|n-m|>1$. It can be convenient to rewrite the Hamiltonian in terms of operators $\hat{S}^j_{n\sigma,m\sigma'}=\hat{c}^{\dag}_{jn,\sigma}\hat{c}_{jm,\sigma'}$, resulting in the following form:
\begin{equation}
\hat{H}/\hbar=\sum_{jn\sigma}n\omega_T\hat{S}^{j}_{n\sigma,n\sigma}+\sum_{jn}\varepsilon_{jn}\hat{S}^{j}_{n\uparrow,n\uparrow}+g_c\sum_{jnm}\zeta_j^{nm}(\hat{S}^{j}_{n\uparrow,m\downarrow}\hat{a}+\hat{a}^{\dag}\hat{S}^{j}_{m\downarrow,n\uparrow})+\delta_c\hat{a}^{\dag}\hat{a}.
\end{equation}
 
In addition to the Hamiltonian dynamics, we also consider dissipation processes such as cavity loss with a rate $\kappa/2\pi=153$~kHz, as well as spontaneous emission with a rate $\gamma/2\pi=7.5$~kHz. The full dynamics of this open system can be described by the following Lindblad master equation:
\begin{equation}
    \frac{\mathrm{d}}{\mathrm{d}t}\hat{\rho}=-\frac{i}{\hbar}[\hat{H},\hat{\rho}]+\bigg[\hat{L}_{\mathrm{cav}}\hat{\rho} \hat{L}_{\mathrm{cav}}^{\dag}-\frac{1}{2}\{\hat{L}_{\mathrm{cav}}^{\dag}\hat{L}_{\mathrm{cav}},\hat{\rho}\}\bigg]+\sum_{jn}\bigg[\hat{L}_{j,n}\hat{\rho} \hat{L}_{j,n}^{\dag}-\frac{1}{2}\{\hat{L}_{j,n}^{\dag}\hat{L}_{j,n},\hat{\rho}\}\bigg],
\end{equation}
where the jump operator for cavity loss is given by $\hat{L}_{\mathrm{cav}}=\sqrt{\kappa}\hat{a}$, and the single-particle jump operators for spontaneous emission are given by $\hat{L}_{j,n}=\sqrt{\gamma}\hat{S}^j_{n\downarrow,n\uparrow}$.
Here, we assume that spontaneous emission is in the Lamb-Dicke regime.
%since the spontaneous emission rate $\gamma$ is much smaller than the axial trapping frequency $\omega_T$.

In the experiment, $\delta_c$ is the largest frequency scale ($\delta_c \gg g_c\sqrt{N},\kappa$), so we can adiabatically eliminate the cavity photons \cite{reiter2012} and obtain the following effective atom-only master equation:
\begin{equation}
    \frac{\mathrm{d}}{\mathrm{d}t}\hat{\rho}=-\frac{i}{\hbar}[\hat{H}_{\mathrm{eff}},\hat{\rho}]+\bigg[\hat{L}_{\mathrm{col}}\hat{\rho} \hat{L}_{\mathrm{col}}^{\dag}-\frac{1}{2}\{\hat{L}_{\mathrm{col}}^{\dag}\hat{L}_{\mathrm{col}},\hat{\rho}\}\bigg]+\sum_{jn}\bigg[\hat{L}_{j,n}\hat{\rho} \hat{L}_{j,n}^{\dag}-\frac{1}{2}\{\hat{L}_{j,n}^{\dag}\hat{L}_{j,n},\hat{\rho}\}\bigg].
\end{equation}
Here, the effective Hamiltonian is given by
\begin{equation}
\hat{H}_{\mathrm{eff}}/\hbar=\sum_{jn\sigma}n\omega_T\hat{S}^{j}_{n\sigma,n\sigma}+\sum_{jn}\varepsilon_{jn}\hat{S}^{j}_{n\uparrow,n\uparrow}+\chi\sum_{jnm}\sum_{kpq}\zeta_{j}^{nm}\zeta_k^{pq}\hat{S}^{j}_{n\uparrow,m\downarrow}\hat{S}^{k}_{p\downarrow,q\uparrow},
\end{equation}
and effective collective jump operator generating superradiant decay takes the form
\begin{equation}
    \hat{L}_{\mathrm{col}}=\sqrt{\Gamma}\sum_{jnm}\zeta_j^{nm}\hat{S}^j_{m\downarrow,n\uparrow},
\end{equation}
where $\chi=-g_c^2\delta_c/(\delta_c^2+\kappa^2/4)$ and $\Gamma=g_c^2\kappa/(\delta_c^2+\kappa^2/4)$. The equivalent superconducting order parameter takes the following form:
\begin{equation}
    \Delta_{\mathrm{BCS}}=\chi\sum_{kpq}\zeta_k^{pq}\langle\hat{S}^k_{p\downarrow,q\uparrow}\rangle.
\end{equation}
One can recover the inhomogeneous model discussed in the previous section by removing the axial harmonic oscillator level labels.

Similarly, the Hamiltonian for initial state preparation takes the form
\begin{equation}
    \hat{H}_{\mathrm{drive}}/\hbar=\sum_{jn\sigma}n\omega_T\hat{S}^{j}_{n\sigma,n\sigma}+\frac{1}{2}\sum_{jnm}\zeta_j^{nm}(\Omega \hat{S}^{j}_{n\uparrow,m\downarrow}+\Omega^{*}\hat{S}^{j}_{m\downarrow,n\uparrow}).
\end{equation}

\begin{figure*}
  \includegraphics[keepaspectratio, width=170mm]{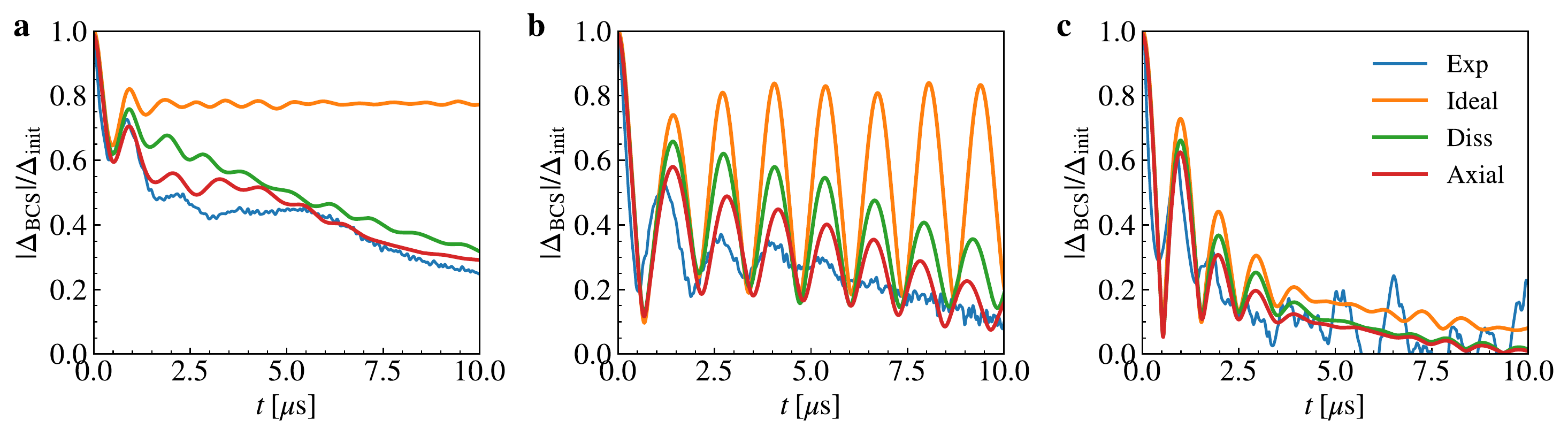}
  \caption{\textbf{Understanding experimental results with axial motion effects.} \textbf{a} Example phase II traces with $\chi N/2\pi=1.29$MHz, $f_{\mathrm{AC}}/2\pi=1.1$MHz. \textbf{b} Example phase III traces with $\chi N/2\pi=0.79$MHz, $f_{\mathrm{AC}}/2\pi=1.1$MHz. \textbf{c} Example phase I traces with $\chi N/2\pi=0.15$MHz, $f_{\mathrm{AC}}/2\pi=1.1$MHz. The blue points are experimental data, the orange lines represent numerical simulations under ideal conditions (see Eq.~(3) in the Methods), the green lines include dissipative processes on top of the ideal simulations, and the red lines consider both dissipative processes and axial motion effects.}
  \label{supp4}
\end{figure*}

In numerical simulations, we perform a mean-field approximation, which replaces the operators $\hat{S}^j_{p\sigma,q\sigma'}$ by their expectation values $\langle\hat{S}^j_{p\sigma,q\sigma'}\rangle$ in the Heisenberg equation of motion. We perform a random sampling of the axial harmonic oscillator mode $n$ for each atom based on a thermal distribution of $15~\mu$K, and we only include the modes $n$ and $n\pm 1$ into our calculation due to the Lamb-Dicke parameter. The atom number in our simulations is set to $2000$; to match $\chi N$ to experimental values, we rescale $\chi$ accordingly. We also empirically take into account two additional dissipation processes to quantitatively capture the behavior of $|\Delta_{\mathrm{BCS}}|$ at longer time scales. The first is a single-particle decoherence between electronic states, described by the jump operators $\hat{L}^{\mathrm{el}}_{j,\sigma}=\sqrt{\gamma_{\mathrm{el}}}\sum_n \hat{S}^j_{n\sigma,n\sigma}$ with $\gamma_{\mathrm{el}}/2\pi<1$kHz for Fig.~2 starting from $t=0\mu$s, and by $\gamma_{\mathrm{el}}/2\pi=0.0036(f_{\mathrm{AC}}/2\pi)+4$kHz for Fig.~3 and Fig.~4 in the main text. The second is a single-particle decoherence between motional states, described by the jump operators $\hat{L}^{\mathrm{mo}}_{j,n}=\sqrt{\gamma_{\mathrm{mo}}}\sum_{\sigma}\hat{S}^j_{n\sigma,n\sigma}$ with $\gamma_{\mathrm{mo}}/2\pi=15$kHz. 

Some example traces including axial motion effects are depicted in Fig.~\ref{supp4}. Generally speaking, accounting for these effects allows us to more accurately predict features present in the experimentally measured evolution of $|\Delta_{\mathrm{BCS}}|$, at the same time leaving the predicted dynamical phase boundaries unchanged. As shown in Fig.~\ref{supp4}a, including axial motion effects in phase II traces allows us to capture the faster damping rate of the Higgs oscillations, as well as a slow oscillation in $|\Delta_{\mathrm{BCS}}|$ at the axial trapping frequency.
Likewise, as shown in Fig.~\ref{supp4}b, including axial motion effects in phase III traces allows us to capture the faster damping rate of the oscillations in $|\Delta_{\mathrm{BCS}}|$, although the observed damping rate is still faster than the rate predicted by theory.
Finally, as shown in Fig.~\ref{supp4}c, all the theory simulations of phase I dynamics are similar to the simulation under ideal conditions, indicating that axial motion does not play an significant role in this regime.

\end{document}